\documentclass[draftclsnofoot, onecolumn, 12pt]{IEEEtran}
\usepackage{graphicx}
\usepackage{amsmath}
\usepackage{times}
\usepackage{latexsym}
\usepackage{graphicx}
\usepackage{bm}
\usepackage{bbm}
\usepackage{amssymb}
\usepackage{stfloats}
\usepackage{cases}
\usepackage{array}
\usepackage{setspace}
\usepackage{fancyhdr}
\usepackage{color}
\usepackage{subfigure}
\usepackage[square, comma, sort&compress, numbers]{natbib}

\usepackage{algorithm}
\usepackage{algorithmicx}
\usepackage{algpseudocode}
\usepackage{amsthm}
\usepackage{multirow}
 \usepackage{mathrsfs}

\def\bm#1{\mbox{\boldmath $#1$}}

\newtheorem{theorem}{Theorem}

\newtheorem{definition}{Definition}

\newtheorem{remark}{Remark}

\newtheorem{problem}{Problem}
\allowdisplaybreaks

\providecommand{\keywords}[1]{\textbf{\textit{Index terms---}} #1}
\begin{document}
\title{ Queue-Aware STAR-RIS Assisted NOMA Communication Systems}
\author{{Nannan Zhang,~\IEEEmembership{Graduate Student Member,~IEEE}, Yuanwei Liu,~\IEEEmembership{Senior Member,~IEEE}, Xidong Mu,~\IEEEmembership{Graduate Student Member,~IEEE}, \\ Wei~Wang,~\IEEEmembership{Senior Member,~IEEE}}
\thanks{N. Zhang and W. Wang are with College of Information Science and Electronic Engineering, Zhejiang University, Hangzhou 310027, China (e-mail: zhangnannan@zju.edu.cn; wangw@zju.edu.cn).}
\thanks{Y. Liu is with the School of Electronic Engineering and Computer
		Science, Queen Mary University of London, London E1 4NS, U.K. (e-mail: yuanwei.liu@qmul.ac.uk).}
\thanks {X. Mu is with the School of Artificial Intelligence, Beijing University of Posts and Telecommunications, Beijing 100876, China (e-mail:
	muxidong@bupt.edu.cn).}}
\maketitle
\vspace{-1.55em}
\begin{abstract}
Simultaneously transmitting and reflecting reconfigurable intelligent surfaces (STAR-RISs) have been receiving great attention nowadays due to the capability of achieving full-space coverage. 
In this paper, the queue-aware STAR-RIS assisted non-orthogonal multiple access (NOMA) communication system is investigated to ensure the system stability. \textcolor{black}{To tackle the challenge of infinite time periods required for stability, 
the long-term stability-oriented problem is reformulated as a queue-weighted sum rate (QWSR) maximization problem in each single time slot based on the Lyapunov drift theory.} In particular, the rate weight allocated to each user is determined by the length of a data queue, which is maintained at the base station (BS) and pending delivery to each user. Then, the QWSR is maximized by jointly optimizing the NOMA decoding order, the active beamforming coefficients (ABCs) at the BS, and the passive transmission and reflection coefficients (PTRCs) at the STAR-RIS, where three STAR-RIS operating protocols are considered, namely energy splitting (ES), mode switching (MS), and time switching (TS). For ES, to handle the highly-coupled and non-convex problem, the blocked coordinate descent and the successive convex approximation methods are invoked to iteratively and alternatively optimize the problem. 
Moreover, the proposed iterative algorithm is further extended to a penalty-based two-loop algorithm to solve the binary amplitude constrained problem for MS. For TS, the formulated problem is decomposed into two subproblems, each of which can be solved in a similar manner as introduced for ES. Simulation results show that: i) our proposed STAR-RIS assisted NOMA communication achieves better performance compared with the conventional schemes; ii) the reformulated QWSR maximization problem is proven to ensure the system stability; and iii) the TS protocol achieves superior performance with respect to both the QWSR and the average queue length. 
\end{abstract}


\section{Introduction}
Reconfigurable intelligent surfaces (RISs)~\cite{RISsurvey2021,RISsurvey2020}, which are also known as intelligent reflecting surfaces (IRSs)~\cite{ZhangR2021,ZhangR2019}, have emerged as a promising and effective technology in the development of wireless communications.
Connected with intelligent controllers (e.g., field-programmable gate array (FPGA)), the two-dimensional (2D) RIS is able to adaptively adjust the phase and even the amplitude of the incident signals, through changing the responses of the reconfigurable elements on the surface, so as to reconfigure the propagation of incident wireless signals and realize the smart radio environment (SRE)~\cite{SRE2020}.
Compared to the conventional relaying technologies, RISs are more economical and environmentally friendly due to the fact that no radio frequency chains are required to manipulate the radio waves and control the propagation of signals.

However, the conventional RIS in most of the existing studies is reflecting-only~\cite{half2019,half2020,half2021}, which leads to the half-space coverage. To tackle this problem, a new technique called simultaneously transmitting and reflecting RIS (STAR-RIS)~\cite{Full2020,WuCoverage,Mu2021,NiSTAR} has received extensive attention from both academia and industry. 
\textcolor{black}{With STAR-RISs, the wireless signals are not only reflected into the same side of the incident signals, but also transmitted into the opposite side.} 
As a result, the STAR-RISs extend the half-space coverage of the conventional reflecting-only RIS into the full-space coverage.
The unique differences between the STAR-RISs and conventional reflecting-only RISs have been discussed~in~\cite{LiuSTAR360} from the perspectives of the hardware design, physics principles, and communication system design.

On the other hand, the non-orthogonal multiple access (NOMA) technique has been already proposed for the third-generation partnership projects long-term evolution advanced (3GPP-LTE-A)~\cite{3Gpp}. It constitutes a promising technology for addressing the large-scale access challenges in 5G and beyond networks by allowing several users to access the wireless network within the same orthogonal resource block (RB)~\cite{Liu2017}. By doing this, more significant bandwidth efficiency can be enhanced than the conventional orthogonal multiple-access (OMA) techniques~\cite{Dai2018}. The core idea of power-domain NOMA~\cite{PowerNOMA2020} is to ensure that multiple users with different power levels can be served within a given time/frequency RB, by employing superposition coding (SC) techniques at the transmitter and the successive interference cancellation (SIC) at the receiver. 
The employed NOMA scheme will achieve more significant gains if the paired users have distinct channel conditions~\cite{Bariah2021}.
Fortunately, the deployment of the STAR-RIS makes it possible to adjust the channel gains among different NOMA users.
Thus, by leveraging the potential benefits of effectively integrating the STAR-RIS with NOMA-assisted communications in this paper, we could explore the performance improvement offered by the aforementioned advantages of these two technologies. 

\subsection{Related works}

\subsubsection{RIS-enabled NOMA communications}
There have been multiple studies focused on conventional reflecting-only RIS-enabled NOMA communications.
In~\cite{Ding2020}, a simple design of RIS assisted NOMA downlink transmission was proposed, where conventional spatial division multiple access was used at the base station (BS) to generate orthogonal beams, and the RIS-assisted NOMA ensures that additional cell-edge users can also be served on these beams.
In~\cite{YangTWC2021}, a combined-channel-strength based user-ordering scheme for NOMA decoding was first proposed to optimize the rate performance in the RIS-assisted downlink NOMA system. Some authors focused on the multiple-input single-output (MISO) scenario with the assistance of RIS and NOMA techniques. 
In~\cite{MuTWC2020}, the RIS-aided MISO NOMA system was investigated to maximize the sum rate both for the ideal and the non-ideal RIS cases, by jointly optimizing the active and passive beamforming vectors. 
In~\cite{HuTCOM2021}, multiple RISs and primary receivers (PRs) were involved, where each RIS acted as an IoT device transmitting information to a corresponding PR to maximize the weighted sum-rate (WSR) of both the primary and the IoT transmissions given the transmit power constraint.
\textcolor{black}{In~\cite{ZhuTCOM2021}, the authors compared the RIS-assisted NOMA scheme and the RIS-assisted zero-forcing beamforming (ZFBF) transmission scheme and then identified the best scenarios to adopt NOMA or ZFBF. 
 In contrast to the alternating optimization techniques, the authors in~\cite{learning2022} conceived a novel smart reconfigurable terahertz (THz)
multiple-input multiple-output (MIMO)-NOMA framework, 
where a novel multi-agent deep reinforcement learning algorithm was proposed by exploiting the decentralized partially-observable Markov decision process. 
As a step further, in~\cite{DLRL2022}, both a deep learning approach and a reinforcement learning approach were developed for the RIS-Assisted NOMA networks to maximize the effective throughput of the entire transmission period. 
Furthermore, instead of signal enhancement, a signal cancellation based design was proposed in~\cite{Hou2020} for the passive beamforming weight at RISs in a MIMO assisted NOMA network.}

\subsubsection{STAR-RIS assisted communications}
\textcolor{black}{Motivated by the full covarage of STAR-RISs, employing STAR-RISs into wireless communications has attracted some initial research interest.}
As stated in~\cite{Mu2021,Niu2021Letters}, there are three STAR-RIS operating protocols including the energy splitting (ES) protocol, mode splitting (MS) protocol , and the time switching (TS) protocol. In~\cite{Mu2021}, a penalty-based iterative algorithm was proposed for the ES protocol of the STAR-RIS under a two-user downlink MISO scenario, which was also extended to the MS protocol.  
In~\cite{Niu2021Letters}, a path-following based technique was first developed to handle the non-convex problem and design the beamforming and the transmitting and reflecting coefficients (TARCs) in an alternating manner.
Moreover, the integration of STAR-RISs and NOMA has also drawn much attention from researchers. 
For example, both NOMA and OMA cases incorporating the STAR-RIS technology were considered in~\cite{WuCoverage, Chenyu2021}. In~\cite{WuCoverage}, a two-user sum coverage range maximization problem was formulated for both NOMA and OMA to optimize both the resource allocation at the access point and the transmission and reflection coefficients at the STAR-RIS. In~\cite{Chenyu2021}, the resource allocation problem in a STAR-RIS-assisted multi-carrier communication network was investigated to maximize the system sum rate.
As a further development, the authors in~\cite{NiSTAR} integrated NOMA and over-the-air federated learning into a unified framework using one STAR-RIS to overcome the spectrum scarcity and support different services.
Moreover, the authors in~\cite{XuSTAR} investigated the hardware model and the channel model for the near-field and the far-field scenarios for STAR-RISs.
Four practical hardware implementations of STAR-RISs, as well as three hardware modeling techniques and five channel modeling methods were discussed in~\cite{XuIOS}.

\subsection{Motivation and Contributions}
\textcolor{black}{Note that most existing research focused on sum rate maximization problems~\cite{MuTWC2020,Chenyu2021} or the constant WSR maximization problems~\cite{HuTCOM2021,Niu2021Letters}, and ignored the system stability issues. 
However, in practice scenarios, there is often a bursty traffic pattern for data services (e.g., video sources)~\cite{Brusty,BrustyData}, which leads to the data accumulation in a queue. 
Unstable queues will cause infinite delays for the users to get their required data. Thus, it is of vital importance to concentrate on the stability of the queueing system~\cite{Qrate2018}, which requires that the data queue length does not go to infinity over a long period of time.
However, the aforementioned studies cannot be applied to the queueing scenarios since they may cause that some queues grow unexpectedly infinitely while other queues are always empty. This consequence reflects the unreasonable allocation and underutilization of wireless resources. 
In fact, both NOMA and STAR-RIS technologies are supposed to help serve more users and maximize the utilization of scarce resources.} To the best of our knowledge, there are no efforts devoted to the system stability problem when employing the STAR-RIS into wireless communications.  
The stability-oriented STAR-RIS assisted NOMA communication problem is non-trivial to be solved owing to the following challenges.
	 \begin{itemize}
	 	\item First, for the considered stability-oriented problem, it requires that the communication system evolves over infinite time periods, and the adjacent time periods are coupled together by the changes in data queues, hence imposing a high level of difficulty.
	 	\item Second, due to the introduction of the transmission coefficients, the optimization of the STAR-RIS-aided system becomes much more challenging than that of the conventional reflecting-only system. The difficulty lies in that STAR-RIS requires optimizing both passive transmission and reflecting beamforming coefficients, which are coupled together by energy conservation. Resource allocation is therefore further complicated. 
	 	\item Third, the NOMA decoding order among the users is determined by not only the active beamforming coefficients (ABCs) at the BS, but also the passive transmission and reflection coefficients (PTRCs) at the STAR-RIS, which leads to a highly-coupled problem. 
	 \end{itemize}
 
\textcolor{black}{Against the above challenges, we jointly investigate the ABCs at the BS and the PTRCs at the STAR-RIS to stabilize the considered queueing communication system.} The main contributions of this paper are summarized as follows.
\begin{itemize}
	\item We consider a queue-aware STAR-RIS assisted downlink MISO-NOMA communication system, where a couple of data queues maintained at the BS are pending to be sent to the users via the STAR-RIS-aided transmission and reflection links. \textcolor{black}{To deal with the challenge of the infinite time duration involved in the system stability, we reformulate the \textcolor{black}{long-term stability-oriented} problem to maximize the per-slot queue-weighted sum rate of users via jointly optimizing the active beamforming and passive beamforming in each single time slot, where three operating protocols of the STAR-RIS, namely ES, MS, and TS, are considered.}
	\item For the ES protocol, to handle the intrinsically coupled non-convex problem, we explore the blocked coordinate descent (BCD) method and the successive convex approximation (SCA) method to iteratively optimize the lower bound of the original problem. 
	We also prove that the rank of the obtained active beamforming vector always satisfies the rank-one constraint.
	\item For the MS protocol, we extend the algorithm for ES into a two-loop penalty-based iterative algorithm \textcolor{black}{to deal with the binary amplitude constrained problem.} For the TS protocol, we decompose the optimization problem into two subproblems, each of which can be solved with the same method adopted for ES.
	\item The numerical results reveal that: i) the STAR-RIS assisted NOMA communications outperform the conventional reflection-only RIS-enabled NOMA communications and the STAR-RIS enabled OMA communications, which verifies the effectiveness of integrating the STAR-RIS with NOMA techniques; ii) the reformulated QWSR maximization problem is able to ensure the system stability; and iii) among the three protocols, the TS protocol achieves superior performance with respect to both the QWSR and the average queue length performance. 
\end{itemize}

\subsection{Organizations}
The rest of this paper is organized as follows. 
The system model and problem formulation are demonstrated in Section~\ref{Sec_model}. Then the Lyapunov drift based stability-driven optimization problem is reformulated in Section~\ref{sec_reformulation}. In Section~\ref{sec_ESsolution}, an efficient algorithm is developed for determining the active beamforming and passive transmission and reflection coefficients for ES. Section~\ref{sec_MSTSsolution} extends the proposed solution for the MS and TS protocols. Following this, the simulation results are provided in Section~\ref{sec_simulation}. Finally, this paper is concluded in Section~\ref{sec_conclusion}.

Notations: Scalars, vectors, and matrices are denoted by lower-case letters, bold-face lower-case letters, and upper-case letters, respectively. Real-valued and complex-valued matrices with the dimension of $N\times M$ are denoted by $\mathbb{R}^{N\times M}$ and $\mathbb{C}^{N\times M}$, respectively. $\textbf{I}_N$ is an $N\times N$ identity matrix. The rank and the trace of matrix $\bm{\mathrm{A}}$ are denoted by $\rm{Rank}(\textbf{A})$ and $\mathrm{Tr}(\bm{\mathrm{A}})$. The diagonal elements of matrix $\bm{\mathrm{A}}$ are denoted by $\rm{Diag}(\bm{\mathrm{A}})$. The positive semidefinite matrix $\bm{\mathrm{A}}$ is represented by $\bm{\mathrm{A}}\succeq 0$. Besides, $\bm{a}^T$, $\bm{a}^H$, and $\rm{diag}(\bm{a})$ denote the transpose, the conjugate transpose, and the diagonal matrix of vector $\bm{a}$, respectively. 

\section{System Model and Problem Formulation}\label{Sec_model}

\begin{figure}[tb]
	\centering
	\includegraphics[width =4 in]{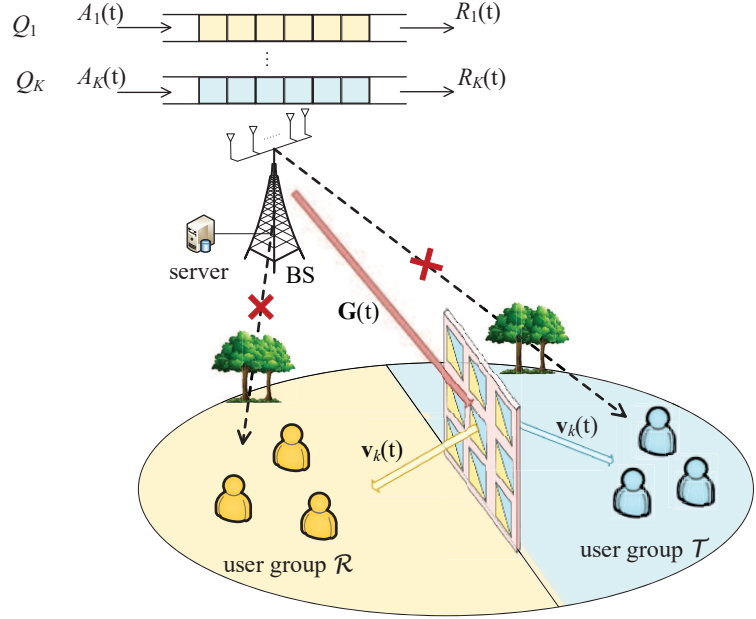}
	\caption{Queue-aware STAR-RIS assisted NOMA communication system.}
	\label{Fig.model}
\end{figure}

We consider a STAR-RIS assisted downlink communication scenario, where an $N$-antenna BS is sending data to $K$ single-antenna users, with the aid of an $M$-element STAR-RIS. \textcolor{black}{A data server is installed on the BS to buffer the pending-transmission data.} 
Let $\mathcal{M}=\{1,2,\cdots,M\}$ denote the set of the STAR-RIS elements. All the users are denoted by the set $\mathcal{K}=\{1,2,\cdots,K\}$. Meanwhile, the users are separated into two groups according to their locations. \textcolor{black}{Denote the group located on the one side of the STAR-RIS by the reflection set $\mathcal{R}$, and denote the other group located on the opposite side of the STAR-RIS by the transmission set $\mathcal{T}$. }  
Assume that the direct communication between the BS and the users is blocked by obstacles, such as the trees and the buildings. This is one of the most typical scenarios in which the STAR-RIS is employed in conventional communication systems~\cite{LiuSTAR360}. Let $t$ denote the time index. The channel from the BS to the STAR-RIS is denoted by $\bm{\mathrm{G}}(t)\in\mathbb{C}^{M\times N}$, while
the channel from the STAR-RIS to the users is denoted by $\textbf{v}_k(t) \in\mathbb{C}^{1\times M}, \forall k\in \mathcal{K}$. 
Suppose that all these channels are narrow-band quasi-static fading. Here, all channel state information is assumed to be acquired perfectly via the channel estimation techniques proposed in~\cite{CSI2020}.

\subsection{Signal model of the STAR-RIS}
For the STAR-RIS, let
$\bm{\Theta}_r(t)=\mathrm{ diag}(\sqrt{\beta^r_1(t)}e^{j\theta^r_1(t)},\sqrt{\beta^r_2(t)}e^{j\theta^r_2(t)},\dots,\sqrt{\beta^r_M(t)}e^{j\theta^r_M(t)})$ as the reflection-coefficient matrix,
and $\bm{\Theta}_t(t)=\mathrm{diag}(\sqrt{\beta^t_1(t)}e^{j\theta^t_1(t)},\sqrt{\beta^t_2(t)}e^{j\theta^t_2(t)},\dots,\sqrt{\beta^t_M(t)}e^{j\theta^t_M(t)})$ as the transmission-coefficient matrix.
\textcolor{black}{Both the amplitude and the phase shift of each STAR-RIS element are assumed to be adjusted continuously, that is, 
$\theta^s_m(t)\in [0,2\pi)$, $\beta_m^s(t)\in[0,1]$,$\forall s \in\{r,t\}, m\in\mathcal{M}$.
Then considering the law of energy conservation, it should be satisfied that $\beta^t_m(t)+\beta^r_m(t)=1$.}

In addition,  
\textcolor{black}{according to the different specific constraints on the amplitude of each STRA-RIS element, there are three protocols for operating the STAR-RIS in wireless communication systems~\cite{LiuSTAR360,Mu2021},} which are listed as follows: 

\begin{enumerate}
\item Energy Splitting (ES) protocol,
	where all elements operate in the simultaneous reflection and transmission mode, that is, \textcolor{black}{the feasible set for the amplitude coefficients $\mathcal{F}^{\rm{ES}}_{\beta}=\{\beta^t_m(t),\beta^r_m(t):\beta^t_m(t),\beta^r_m(t)\in[0,1],\beta^t_m(t)+\beta^r_m(t)=1\}$.} 
\item Mode Switching (MS) protocol,
	where some of the elements operate in the full reflection mode, while the other elements operate in the full transmission mode, namely \textcolor{black}{$\mathcal{F}^{\rm{MS}}_{\beta}=\{\beta^t_m(t),\beta^r_m(t):\beta^t_m(t),\beta^r_m(t)\in\{0,1\},\beta^t_m(t)+\beta^r_m(t)=1\}$}. 
\item Time Switching (TS) protocol,
	where all the elements are switched to operate in the full reflection mode and the full transmission mode 
periodically in different time periods, denoted by the R period and T period. 
\textcolor{black}{Denote the feasible amplitude set of the R period by $\mathcal{B}_{\beta}^R=\{\beta^t_m(t),\beta^r_m(t):\beta^r_m(t)=1, \beta^t_m(t)=0, \forall m \in \mathcal{M}$\}, 
and that of the T period by  $\mathcal{B}_{\beta}^T=\{\beta^t_m(t),\beta^r_m(t):\beta^r_m(t)=0, \beta^t_m(t)=1, \forall m \in \mathcal{M} \} $. 
Then for TS, we have $\mathcal{F}^{\rm{TS}}_{\beta}=\{\beta^t_m(t),\beta^r_m(t):\beta^t_m(t),\beta^r_m(t)\in \mathcal{B}_{\beta}^R \cup \mathcal{B}_{\beta}^T \}$}. But as a result, there are other constraints on the time percentage allocated to the R period, $\alpha^r(t)$ and to the T period, $\alpha^t(t)$, shown as $\alpha^r(t)+\alpha^t(t)=1$, $\alpha^r(t),\alpha^t(t)\in[0,1]$. 
\end{enumerate}

\subsection{STAR-RIS assisted NOMA communication Model}
Let $\textbf{w}_{k}(t)\in\mathbb{C}^{N\times 1}$ and $x_{k}(t)$ denote the active beamforming vector and the information-bearing symbol for user $k\in\mathcal{K}$ at the BS, respectively. Let $P_{max}$ denote the maximum transmitted power of the BS, then we have
\begin{equation}
\sum_{k\in\mathcal{K}}\textbf{w}^H_{k}(t)\textbf{w}_{k}(t)\leq P_{max}.
\end{equation} 

\subsubsection{\textcolor{black}{For ES and MS}} Suppose that all users are grouped together to form the NOMA pairs. The received signal at user $k\in\mathcal{K}$ is
\vspace{-0.5 em}
\begin{spacing}{1}
\begin{equation}
y_k(t)=\textbf{v}_k(t)\bm{\Theta}_{s_k}(t)\textbf{G}(t)\sum_{i\in\mathcal{K}} \textbf{w}_{i}(t)x_{i}(t)+n_k(t),
\end{equation}
\end{spacing}
\noindent \textcolor{black}{where $s_k\in\{r,t\}$ indicates one of the half spaces of the STAR-RIS where user $k$ is located, and $s_k=r \text{ if } k \in \mathcal{R}$ while $s_k=t \text{ if } k \in \mathcal{T}$, $\mathbb{E}[|x_{i}(t)|^2]=1$,} and $n_k(t)\in\mathcal{CN}(0,\sigma^2)$ is the additive white Gaussian noise at user $k$.

To eliminate the interference efficiently, SIC is utilized at each user according to the NOMA principle. Let $o_k(t)$ denote the decoding order of user $k\in \mathcal{K}$.
The smaller the decoding order $o_k(t)$ is, the earlier its signal is decoded, and the more interference this user will suffer.
By treating the signals of users with a larger decoding order as interference, 
the achievable signal-to-interference-plus-noise ratio (SINR) for user $k$ decoding its own signal is
\vspace{-0.5 em}
\begin{spacing}{1.2}
\begin{equation}
\mathrm{SINR}_{kk}(t)=\frac{|\textbf{v}_k(t)\bm{\Theta}_{s_k}(t)\bm{\textbf{G}}(t)\textbf{w}_k(t)|^2}{\sum_{i:o_k(t)<o_i(t)}|\textbf{v}_k(t)\bm{\Theta}_{s_k}(t)\bm{\textbf{G}}(t)\textbf{w}_{i}(t)|^2+\sigma^2}.
\end{equation}
\end{spacing}
\noindent  
Moreover, the SINR achieved by decoding the signal of user $k$ at user $j$, the one with a larger decoding order (i.e., $o_k(t)<o_j(t)$), is
\vspace{-0.5 em}
\begin{spacing}{1.2}
\begin{equation}
\mathrm{SINR}_{kj}(t)=\frac{|\textbf{v}_{j}(t)\bm{\Theta}_{{s_j}}(t)\bm{\textbf{G}}(t)\textbf{w}_k(t)|^2}{\sum_{i:o_k(t)<o_i(t)} |\textbf{v}_{j}(t)\bm{\Theta}_{{s_j}}(t)\bm{\textbf{G}}(t)\textbf{w}_{i}(t)|^2\!+\!\sigma^2},\text{if } o_k(t)\!<\!o_j(t), \forall i \!\in \!\mathcal{K}.
\end{equation}
\end{spacing}
\noindent

\subsubsection{\textcolor{black}{For TS}} Note that the TS protocol allows the STAR-RIS to operate in the T period or the R period alternatively. Thus, we can adopt the NOMA in different user groups distributed on the different sides of the STAR-RIS within every period. 
Along this line, 
the SINR that user $k$ decodes the signal of itself is
\vspace{-0.8 em}
\begin{spacing}{1.2}
	\begin{equation}
	\mathrm{SINR}_{kk}^s(t)=\frac{|\textbf{v}_k(t)\bm{\Theta}_s(t)\textbf{G}(t)\textbf{w}_k(t)|^2}{\sum_{i:o_k(t)<o_i(t)}|\textbf{v}_k(t)\bm{\Theta}_s(t)\textbf{G}(t)\textbf{w}_{i}(t)|^2+\sigma^2},   s=\bigg\{\begin{array}{ll}
	r, \forall k,i \in \mathcal{R}\\ t, \forall k,i \in \mathcal{T}
	\end{array}.
	\end{equation}
\end{spacing}
\noindent 
 Additionally, the SINR of the signal of user $k$ decoded at user $j$ is
	\vspace{-1.3 em}
	\begin{spacing}{1.2}
	\begin{equation}
	\mathrm{SINR}_{kj}^s(t)\!=\!\frac{|\textbf{v}_{j}(t)\bm{\Theta}_{s}(t)\textbf{G}(t)\textbf{w}_k(t)|^2}{\sum_{i:o_k(t)<o_i(t)} |\textbf{v}_{j}(t)\bm{\Theta}_{s}(t)\textbf{G}(t)\textbf{w}_{i}(t)|^2\!+\!\sigma^2}, \text{if }o_k(t)\!<\!o_j(t); s\!=\!\bigg\{\begin{array}{ll}
	\!	r, \forall k, j,i \!\in \!\mathcal{R}\\\! t, \forall k,j,i \!\in \!\mathcal{T}
	\end{array}\!.
	\end{equation}
\end{spacing}\noindent
 
Define $R_k$ as the achievable rate of user $k$ decoding its own signal.
For successful SIC operations, it is crucial that $R_k$ is limited by the minimum of the achievable rates at which user $j$ as well as user $k$ can decode the signal of user $k$~\cite{SICsuccess,NomaRate}. 
\textcolor{black}{To this end, the following conditions should be satisfied for the SIC to be applied successfully, given by
\vspace{-0.5em}
\begin{spacing}{1}
\begin{equation}
R_{k}(t) \le \min \big\{\log_2(1+\mathrm{SINR}_{kk}(t)),\log_2(1+\mathrm{SINR}_{kj}(t))\big\}, \text{ if } o_k(t)<o_j(t).
\end{equation}
\end{spacing}
}

In addition, we should guarantee that more wireless resources are allocated to the user with a lower decoding order to keep itself a reasonable communication rate, since this user is suffering more interference than the one with a higher order. Thus, for a given decoding order, the following $K(K-1)$ rate fairness conditions should be satisfied,
\vspace{-0.5em}
\begin{spacing}{1}
\begin{equation}
|\textbf{v}_{i}(t)\bm{\Theta}_{s_i}(t)\bm{\textbf{G}}(t)\textbf{w}_k(t)|^2  \ge |\textbf{v}_{i}(t)\bm{\Theta}_{s_i}(t)\bm{\textbf{G}}(t)\textbf{w}_{j}(t)|^2,\text{if } o_k(t) < o_j(t), \forall i\in \mathcal{K},
\end{equation}
\end{spacing}
\noindent where $s_i= r \text{ if } i \in \mathcal{R}$ and $t, \text{ otherwise}$.

\subsection{Queue Dynamics and Stability}

\textcolor{black}{ Different from the existing studies ignoring the data arrival process, we consider a bursty data arrival at the BS.} Let $\textbf{A}(t) =
\{A_k(t), \forall k\in\mathcal{K}\}$ be the random arrival from the data servers. Assume that $\textbf{A}(t)$ is independent and identically distributed over time slots, where $E[A_k(t)] =\lambda_k$ and $\lambda_k$ is the average arrival rate of the data transmitted to user $k$.
A data queue is maintained at data server of the BS for sending to the corresponding user with the serving rate being $R_{k}(t)$. Let $Q_k(t)$ denote the queue length for user $k$ at the current slot, then we have the following queue dynamics for user $k$ at the next slot:
\vspace{-0.3em}
\begin{spacing}{1.1}
\begin{equation}\label{Qdynamics}
Q_k(t+1)=[Q_k(t)-R_{k}(t)\tau]^+ + A_k(t)\tau, \forall k \in \mathcal{K},
\end{equation}
\end{spacing}
\noindent where $[\cdot]^+$ is equal to the number enclosed in the parenthesis if this number is nonnegative and to 0, otherwise. Meanwhile, $\tau$ is the time duration of one slot.
\begin{definition}{(Queue Stability):}
	A queue $Q_k(t)$ is strongly stable if 
	\begin{equation}\label{Queue_stability}
	\lim_{T\to\infty}\frac{1}{T}\left(\sum_{t=0}^{T}\mathbb{E}\big(Q_k(t)\big)\right)<\infty.
	\end{equation}
\end{definition}
The system is said to be stable if all the queues in the system are strongly stable. To ensure that the system is stabilizable, the capacity region following~\cite{Neely2010} is defined as follows:
\begin{definition}{(Capacity Region):}
	The capacity region 
	is defined as the closure of the set of all input rate vectors
	$\bm{\lambda}\triangleq\{\lambda_k\}$ stabilizable under some rate allocation algorithm.
\end{definition}

\subsection{Problem Formulation}
Our goal is to stabilize the system for any arrival rate vector $\bm{\lambda}$ strictly interior to the capacity region, by jointly optimizing the active beamforming coefficients (ABCs) at the BS, the passive transmission and reflection coefficients (PTRCs) at the STAR-RIS, and the NOMA decoding order. From this viewpoint, this problem can be formulated as
\vspace{-0.85em}
\begin{spacing}{1.1}
\textcolor{black}{\begin{subequations}
	\begin{align}
     \text{find} \!\! \! \quad &  \{\textbf{w}_{k}(t),\bm{\Theta}_{s}(t),o_{k}(t),\alpha^s(t)\}
    \label{find}\\
\rm{s.t. }\quad	& Q_{k}(t), \forall k \in\mathcal{F}^{X}_u \text{ is strongly stable},
\label{Qk}\\
& \sum_{k\in\mathcal{F}^{X}_u}\textbf{w}^H_{k}(t)\textbf{w}_{k}(t)\leq P_{max},
\label{a}\\
	& 
	 \beta^r_m(t), \beta^t_m(t) \in\mathcal{F}^{X}_{\beta},
\label{b}\\
	& \theta^r_m(t),\theta^t_m(t)\in[0,2\pi), 
\label{c}\\
	&R_{k}(t)\! \le \! \min\!  \big\{\! \log_2(1\! +\! \mathrm{SINR}_{kk}(t)),\! \log_2(1\! +\! \mathrm{SINR}_{kj}(t))\! \big\},\!  \text{ if } o_k(t)\! <\! o_j(t), \forall k,j\in\mathcal{F}^{X}_u,
\label{d}\\
	&|\textbf{v}_{i}(t)\bm{\Theta}_{s_i}(t)\textbf{G}(t)\textbf{w}_k(t)|^2  \ge |\textbf{v}_{i}(t)\bm{\Theta}_{s_i}(t)\textbf{G}(t)\textbf{w}_{j}(t)|^2,\!\text{ if } o_k(t)\!<\! o_j(t), \forall i,k,j\!\in\mathcal{F}^{X}_u,
\label{e}\\
	& \alpha^t(t)+\alpha^r(t)=1, \alpha^r(t),\alpha^t(t)\in[0,1],
	\label{alpha}		
	\end{align}
\end{subequations}}
\end{spacing}
\vspace{-0.85em}
\noindent where ${\rm{X}}\in\{\rm{ES,MS,TS}\}$ represents the operating protocols of the STAR-RIS and $\mathcal{F}^{X}_u$ is the corresponding user set. Specifically, $\mathcal{F}^{X}_u=\mathcal{K}$ if ${\rm{X}}\in\{\rm{ES,MS}\}$ and $\mathcal{F}^{X}_u=\mathcal{R}$ or $\mathcal{T}$ according to which period the STAR-RIS is working in when ${\rm{X}}=\rm{TS}$. 
Constraint~(\ref{Qk}) guarantees that all the data queues are strongly stable so as to stabilize the system.
Constraint~(\ref{a}) is the total transmission power limited at the BS. Constraint~(\ref{b}) indicates the restriction for the energy conservation. Constraint~(\ref{c}) is the reflection and transmission phase shift for each element of the STAR-RIS. Moreover, constraint~(\ref{d}) provides the decoding order conditions for SIC, and constraint~(\ref{e}) ensures the rate fairness among users.  Constraint~(\ref{alpha}) is related to the time allocation variables for TS, which is invalid for ES and MS.

\section{stability-driven reformulated optimization}\label{sec_reformulation}
Note that the stability requirement in constraint~(\ref{Qk}) for each queue demands the evolution over an infinite time horizon, which is non-trivial to be solved. To address this problem, we resort to the Lyapunov drift approach~\cite{Neely2010} to divide the queue stability into the effort of minimizing the drift in each time slot so that the stability constraints are met in the long term.

To be more specific, we adopt a widely-used quadratic Lyapunov function~\cite{Cui2012}, which increases quadratically with the queue length, given by
\vspace{-0.5 em}
\begin{spacing}{1.1}
\begin{equation}
L(Q_k(t))=\sum_{k\in\mathcal{K}}\big(Q_k(t)\big)^2,
\end{equation}
\end{spacing}
\noindent and then the Lyapunov drift in slot $t$ is given by
\vspace{-0.5em}
\begin{spacing}{1.1}
\begin{equation}\label{Lydrift}
\Delta\big(Q_k(t)\big)=\mathbb{E}\Big(L\big(Q_k(t+1)-L(Q_k(t)\big)\Big).
\end{equation}
\end{spacing}
\noindent Taking the square operation on both sides of~(\ref{Qdynamics}), we have
\vspace{-0.8em}
\begin{spacing}{1.1}
\begin{equation}\label{Qsquare}
\big(Q_k(t+1))^2\le\big(Q_k(t))^2+\big(R_{k}(t)\big)^2+(A_k(t))^2+2Q_k(k)A_k(t)-2Q_k(t)R_{k}(t).
\end{equation}
\end{spacing}
\noindent Sum over $k$ on both sides of~(\ref{Qsquare}) and rearrange the items, then based on the definition in~(\ref{Lydrift}), we have
\vspace{-0.8em}
\begin{spacing}{1}
 \begin{equation}\label{drift}
 \Delta\big(Q_k(t)\big)\le Bo+2\sum_{k\in\mathcal{K}}{Q_k(k)\lambda_k(t)}-2\sum_{k\in\mathcal{K}}Q_k(t)R_{k}(t),
 \end{equation}
\end{spacing}
\noindent where $Bo$ is a bounded constant, given by $Bo=\sum_k\mathbb{E}\big(\big(R_{max}(t)\big)^2\big)+\sum_k\mathbb{E}\big((A_k(t))^2\big)$ with $R_{max}(t)=\max_k\{R_{k}(t)\}$.
 
To stabilize the system, we minimize the upper bound of the Lyapunov drift (the right hand side (RHS) of~(\ref{drift})) at each time slot~\cite{Qrate2018}, which is equivalent to maximizing the term $\sum_k Q_k(t)R_{k}(t)$, i.e., the queue-weighted sum rate (QWSR) in each time slot. 
In this case, the stability-driven optimization for the STAR-RIS assisted NOMA communication can be casted as the following QWSR-optimal designs.
Note that we omit the time index $t$ in what follows for easier presentation. 
\begin{problem}{(QWSR-optimal STAR-RIS-assisted NOMA Communication Problem)}\label{P0}
\textcolor{black}{	
	\vspace{-0.5 em}
	\begin{spacing}{1}
	\begin{subequations}\label{PESMS}
		\begin{align}
			\max _{\{\boldsymbol{\Theta}_s,\mathrm{\boldsymbol{w}}_k,o_k\}}\quad &  \sum_{k\in\mathcal{K}}Q_kR_{k} \Big/	\max _{\{\boldsymbol{\Theta}_s,\mathrm{\boldsymbol{w}}_k,o_k,\alpha^s\}} \alpha^sQ_kR^s_{k} \label{P0_obj}\\
			\rm{s.t. } \quad &\rm {(\ref{a})-(\ref{alpha})},
		\end{align}
	\end{subequations}
\end{spacing}
\noindent where the parameter $\alpha^s, \forall s \in\{r,t\}$ and the related constraint~(\ref{alpha}) are only valid for the TS protocol.}
\end{problem}
\begin{remark}
It ensures that the queuing system is stable as long as the average arrival rate vector is within the system stability region. Besides, this positive queue length based weight can be regarded as the urgency/priority of the user in resource allocation, which is formed in the media access control (MAC) layer to achieve certain fairness purposes.
\end{remark}

However, the optimization problem~(\ref{PESMS}) is still challenging to be solved directly for the following reasons. First of all, there are multiple highly coupled variables (i.e., $\{\textbf{w}_k\}, \{\bm{\Theta}_s\}$) and the non-convex objective and constraint~(\ref{e}). Furthermore, the decoding order needs to be determined, which is influenced by not only the ABCs at the BS, but also the PTRCs at the STAR-RIS. In the following, we propose an efficient approach to address this challenge.


\section{Proposed Solution for the ES Protocol}\label{sec_ESsolution}
In this section, we focus on solving problem~(\ref{PESMS}) for the ES protocol. \textcolor{black}{For a given SIC decoding order, we first transform the problem into a solvable form. Then, the intrinsically coupled problem is decomposed into two subproblems, 
	which are optimized separately and alternatively.}

For the $K$ users, we can solve the problem \textcolor{black}{$K!$ }times via the exhaustive search scheme, each of which is for one particular user decoding order, to obtain the corresponding performance respectively, and then choose the maximum one. 
Thus, for a given decoding order ${o_k}$, problem~(\ref{PESMS}) for ES can be detailed as 
\vspace{-1em}
\begin{spacing}{1}
\begin{subequations}\label{PES}
	\begin{align}
	\max _{\{\boldsymbol{\Theta}_s,\textbf{w}_k\}} \quad &  \sum_{k\in\mathcal{K}}Q_kR_{k} \label{P1_obj}\\
	\rm{s.t. }  \quad
	& \beta^t_m+\beta^r_m=1,\beta^r_m, \beta^t_m \in[0,1],
\label{1b}\\
& \rm{(\ref{a}), (\ref{c})-(\ref{e})}.
\label{1a}
	\end{align}
\end{subequations}	
\end{spacing}

To facilitate the design, we define the reflection- and transmission- coefficient vectors as $\textbf{d}_s=[\sqrt{\beta^s_1}e^{j\theta^s_1},\sqrt{\beta^s_2}e^{j\theta^s_2},\dots,\sqrt{\beta^s_M}e^{j\theta^s_M}]^H,\forall s\in\{t,r\}$, which means $\bm{\Theta}_{s_k}=\mathrm{diag}(\textbf{d}_{s_k}^H)$. In this case, we have $|\textbf{v}_k\bm{\Theta}_{s_k}\textbf{G}\textbf{w}_k|^2=|\textbf{d}_{s_k}^H\textbf{H}_k\textbf{w}_k|^2$, where $\textbf{H}_k=\mathrm{diag}(\textbf{v}_k)\textbf{G}$. Furthermore, we define $\textbf{D}_s=\textbf{d}_s\textbf{d}_s^H, \forall s\in\{t,r\}$, which satisfies $\textbf{D}_s \succeq \bm{0}$ and $\mathrm{Rank}(\textbf{D}_s)=1$. The diagonal elements of $\textbf{D}_s$ are that $ \mathrm{Diag}(\textbf{D}_s)=\bm{\beta}^s\triangleq[\beta_1^s, \beta_2^s,\dots,\beta_M^s]$. Similarly, let $\textbf{W}_k=\textbf{w}_k\textbf{w}_k^H, \forall k\in\mathcal{K}$, which satisfies $\textbf{W}_k \succeq \bm{0}$ and $\mathrm{Rank}(\textbf{W}_k)=1$. 
Then, by introducing the slack variables $\{S_{kj}\}$ and  $\{I_{kj}\}, \forall k,j\in\mathcal{K}$ as follows
\vspace{-1 em}
\begin{spacing}{1.2}
\begin{align}
\frac{1}{ S_{kj}}&=|\textbf{d}_{s_j}^H\textbf{H}_j\textbf{w}_k|^2
=\mathrm{Tr}(\textbf{W}_k\textbf{H}_{j}^H\textbf{D}_{s_j}\textbf{H}_{j}),\\
I_{kj}&=\sum_{i:o_i>o_k}|\textbf{d}_{s_j}^H\textbf{H}_j\textbf{w}_{i}|^2+\sigma^2
=\sum_{i:o_i>o_k}\mathrm{Tr}(\textbf{W}_{i}\textbf{H}_{j}^H\textbf{D}_{s_j}\textbf{H}_{j})+\sigma^2, 
\end{align}
\end{spacing}
\noindent we can rewrite the SINR as 
\vspace{-1.8 em}
\begin{spacing}{1.2}
\begin{align}\label{RSI}
\mathrm{SINR}_{kj}=\frac{1}{S_{kj}I_{{k}j}},o_k\le o_j.
\end{align}
\end{spacing}
\noindent Till now, problem~(\ref{PES}) for ES can be reformulated as 
\vspace{-0.8 em}
\begin{spacing}{1}
\begin{subequations}\label{PSI}
\begin{align}
	\max _{\substack{\{\textbf{D}_s,\textbf{W}_k,\boldsymbol{\beta}^s,\\R_{k},S_{kj},I_{kj}\} } }\quad &  \sum_{k\in\mathcal{K}}Q_kR_{k} \label{P2_obj}\\
	\rm{s.t. }   \quad
&\frac{1}{ S_{kj}}\le \mathrm{Tr}(\textbf{W}_k\textbf{H}_{j}^H\textbf{D}_{s_j}\textbf{H}_{j}),\text{ if } o_k \le o_{j}, 
\forall k,j\in\mathcal{K},
\label{2b}\\
&I_{kj}\ge \sum_{i:o_i>o_k}\mathrm{Tr}(\textbf{W}_{i}\textbf{H}_{j}^H\textbf{D}_{s_j}\textbf{H}_{j})+\sigma^2, \text{ if } o_k \le o_{j}, 
\forall k,j\in\mathcal{K},
\label{2c}\\
&R_{k}\le \min\Big\{\log_2\big(1+\frac{1}{S_{kk}I_{kk}}\big),\log_2\big(1+\frac{1}{S_{kj}I_{kj}}\big)\Big\}, \text{ if } o_k<o_j,\forall k,j\in\mathcal{K},
\label{2e}\\
& \sum_{k\in\mathcal{K}}\mathrm{Tr}(\textbf{W}_{k})\leq P_{max},
\label{2a}\\
&
 \mathrm{Tr}(\textbf{W}_k\textbf{H}_{i}^H\textbf{D}_{s_i}\textbf{H}_{i})
\ge 
 \mathrm{Tr}(\textbf{W}_j\textbf{H}_{i}^H\textbf{D}_{s_i}\textbf{H}_{i}),
\text{ if } o_k < o_{j}, \forall i\in\mathcal{K},
\label{2d}	\\
& \beta^t_m+\beta^r_m=1,
 \beta^r_m,\beta^t_m\in[0,1],
\label{2f}\\	
&\mathrm{Diag} (\textbf{D}_s)=\bm{\beta}^s,\forall s\in\{t,r\},
\label{2g}\\
&\mathrm{Rank}(\textbf{D}_s)=1,\forall s\in\{t,r\},
\label{2h}	\\
&\mathrm{Rank}(\textbf{W}_k)=1,\forall k\in\mathcal{K},
\label{2i}\\
&\textbf{D}_s \succeq \bm{0},\forall s\in\{t,r\},
\label{2j}\\
&\textbf{W}_k \succeq\bm{0}, k\in\mathcal{K}.
\label{2k}
\end{align}
\end{subequations}
\end{spacing}

\textcolor{black}{ 
In the problem above, the equality of constraints (\ref{2b}) and~(\ref{2c}) can be always guaranteed to make the problems~(\ref{PSI}) and~(\ref{PES}) equivalent. 
When $k=j$, if any strict inequality in constraints~(\ref{2b}) and~(\ref{2c}) holds, we can adjust it by 
decreasing the values of $S_{kj}$ and $I_{kj}$ so that the equality is attained, which increases the value of the objective function.
When $k\ne j$, based on constraint~(\ref{2e}), we can do the same operations but without changing the objective function's value if initially $\log_2\big(1+\frac{1}{S_{kk}I_{kk}}\big)\le \log_2\big(1+\frac{1}{S_{kj}I_{kj}}\big)$, or with increasing the the objective function's value if initially $\log_2\big(1+\frac{1}{S_{kk}I_{kk}}\big)> \log_2\big(1+\frac{1}{S_{kj}I_{kj}}\big)$. As a result, the equivalence between problems~(\ref{PSI}) and~(\ref{PES}) has been established.
}

However, it is still a non-convex optimization problem due to the non-convex objective and constraints~(\ref{2e}),~(\ref{2g})-(\ref{2i}), as well as the non-convex highly coupled terms in constraints~(\ref{2b})-(\ref{2c}) and (\ref{2d}).  In this case, to overcome the challenge, we decompose this problem into two subproblems in terms of the active beamforming optimization at the BS and the passive beamforming optimization at the STAR-RIS, which are optimized separately and iteratively.
\subsection{Active beamforming optimization at the BS}\label{ABC}
For any given passive beamforming coefficients $\{\textbf{D}_s\}$, or saying $\{\textbf{d}_s\},\forall s\in\{t,r\}$, the active beamforming optimization at the BS can be rewritten as
\vspace{-0.8 em}
\begin{spacing}{1.1}
\begin{subequations}\label{PW_ES}
\begin{align}
\max _{\{\textbf{W}_k,R_{k},S_{kj},I_{kj}\}} \quad &  \sum_{k\in\mathcal{K}}Q_kR_{k} \label{P3_obj}\\
\rm{s.t. }   \quad
& \rm{(\ref{2b})-(\ref{2d}),(\ref{2i}),(\ref{2k})}. 
\label{3a}
\end{align}
\end{subequations}
\end{spacing}
\noindent For the non-convex constraint~(\ref{2e}),
it is worth noting that the RHS of constraint~(\ref{2e}) is a joint convex function with respect to $S_{kj}$ and $I_{kj}$ since its Hessian function is semidefinite for any $S_{kj}>0$ and $I_{kj}>0$. If given any local point $\{\tilde{S}_{kj}, \tilde{I}_{kj}\}$, we can get a lower bound at this point by utilizing the first-order Taylor expansion as
\vspace{-0.8 em}
\begin{equation}\label{Taylor}
\log_2\Big(1+\frac{1}{S_{kj}I_{kj}}\Big)\ge R^{low}_{kj}
=\log_2(1+\frac{1}{\tilde{S}_{kj}\tilde{I}_{kj}})
-\frac{S_{kj}-\tilde{S}_{kj}}{\ln 2(\tilde{S}_{kj}^2\tilde{I}_{kj}+\tilde{S}_{kj})}
-\frac{I_{kj}-\tilde{I}_{kj}}{\ln 2(\tilde{I}_{kj}^2\tilde{S}_{kj}+\tilde{I}_{kj})}.
\end{equation}
\noindent
Based on~(\ref{Taylor}), the non-convex constraint~(\ref{2e}) is relaxed to be the following inequality 
\vspace{-0.5 em}
\begin{spacing}{1.1}
\begin{equation}\label{S}
 R_{k}\le \min\{R_{kk}^{low},R_{kj}^{low}\}, \text{ if } o_k < o_{j},
\forall k,j\in\mathcal{K}.
\end{equation}
\end{spacing}
\noindent We reformulate problem~(\ref{PW_ES}) 
which can be approximated as 
\vspace{-0.8 em}
\begin{spacing}{1}
\begin{subequations}\label{PW_ESlow}
\begin{align}
\max _{\{\textbf{W}_k,R_{k},S_{kj},I_{kj}\}} \quad &  \sum_{k\in\mathcal{K}}Q_kR_{k} \label{P33_obj}\\
\rm{s.t. }   \quad
& \rm{(\ref{2b}),(\ref{2c}),(\ref{2a}),(\ref{2d}),(\ref{2i}),(\ref{2k}),
	(\ref{S})}.
\label{33a}
\end{align}
\end{subequations}
\end{spacing}
\noindent
As for the non-convex rank-one constraint~(\ref{2i}), we have the following theorem.
\begin{theorem}\label{Theom_RankW}
	The solution $\{\textbf{W}_k\}$ obtained without the rank-one constraint always satisfies that $\mathrm {Rank}(\textbf{W}_k) = 1, \forall k \in\mathcal{K}$.
\end{theorem}
\begin{IEEEproof}
	Please refer to Appendix~\ref{ProofTheom1}.
\end{IEEEproof}

By exploiting this theorem, we can obtain a rank-one solution by ignoring this constraint directly. Since the final relaxed problem without the rank-one constraint is a standard semidefinite program (SDP)~\cite{Boyd2004}, it can be efficiently solved by well-known convex optimization tools, such as the CVX~\cite{CVX}. The objective value obtained from problem~(\ref{PW_ESlow}) yields a lower bound of that from problem~(\ref{PW_ES}) owing to the relaxation in~(\ref{S}).  
After the solution is derived, we can get the beamforming vector $\{\textbf{w}_k\}$ via Cholesky decomposition as 
$\textbf{W}_k=\textbf{w}_k\textbf{w}_k^H.$

\subsection{Passive beamforming optimization at the STAR-RIS} 
For any given active beamforming vectors 
 $\{\textbf{w}_k\}, \forall k\in\mathcal{K}$, the passive beamforming optimization at the STAR-RIS can be rewritten as
 \vspace{-0.8 em}
 \begin{spacing}{1}
\begin{subequations}\label{PQ_ES}
\begin{align}
\max _{\substack{\{\textbf{D}_s, \boldsymbol{\beta}^s, R_{k},\\S_{kj},I_{kj}\}}} \quad &  \sum_{k\in\mathcal{K}}Q_kR_{k} \label{P6_obj}\\
\rm{s.t. }   \quad
&\rm{(\ref{2b})-(\ref{2e}),(\ref{2d})-(\ref{2h}),(\ref{2j})}.
\label{6a}
\end{align}
\end{subequations}
\end{spacing}

The manipulation for the non-convex constraint~(\ref{2e}) is similar to the methods used in Sec.~\ref{ABC}, from which they be approximated as
~(\ref{S}).
Let $\bm{e}_m$ denote an $M$-dimensional column vector with the $m$-th element being $1$ and all others being $0$. Then, combing constraints~(\ref{2f}) and~(\ref{2g}), we have the equivalent formula below
\vspace{-0.5em}
\begin{spacing}{1.1}
\begin{equation}\label{ee}
\bm{e}_m^H\textbf{D}_r\bm{e}_m+\bm{e}_m^H\textbf{D}_t\bm{e}_m= 1,
\end{equation}
\end{spacing}
\noindent after which the variable $\bm{\beta}^k$ in problem~(\ref{PQ_ES}) is eliminated.
Then the relaxed problem is given by
\begin{spacing}{1}
\begin{subequations}\label{PQES}
\begin{align}
\max _{\{\textbf{D}_s,R_{k} ,S_{kj},I_{kj}\}} \quad &  \sum_{k\in\mathcal{K}}Q_kR_{k}  \label{P7_obj}\\
\rm{s.t. }   \quad
&\rm{(\ref{2b}),(\ref{2c}),(\ref{2d}),(\ref{2h}),(\ref{2j}),
	(\ref{S}),(\ref{ee})}.
\label{7a}
\end{align}
\end{subequations}
\end{spacing}

As for the non-convex rank-one constraint~(\ref{2h}), there is no guarantee that the obtained $\{\textbf{D}_s\}$ by ignoring this constraint can always be rank-one. \textcolor{black}{In this case, the typical SDR approach may fail to work well since the reconstructed rank-one solution via the Gaussian randomization method~\cite{ZhuTCOM2021} may not be feasible. To address this challenge, we turn to find a local optimal rank-one solution by applying the sequential rank-one constraint relaxation (SROCR)-based method~\cite{MuTWC2020,SROCR}, in which the rank-one constraint is replaced with a relaxed convex formula.} 
Then the further relaxed optimization problem is given as
\vspace{-0.5 em}
\begin{spacing}{1}
\begin{subequations}\label{PQES_SROCR}
\begin{align}
\max _{\{\textbf{D}_s,R_{k},S_{kj},I_{kj}\}} \quad &  \sum_{k\in\mathcal{K}}Q_kR_{k}  \label{P71_obj}\\
\rm{s.t. }   \quad
& \textbf{u}_{\max}\big(\textbf{D}_s^{(i)}\big)^H \textbf{D}_s \textbf{u}_{\max}\big(\textbf{D}_s^{(i)}\big)\le \gamma^{i}\rm{Tr}(\textbf{D}_s),
\label{71a}\\
&\rm{(\ref{2b}),(\ref{2c}),(\ref{2d}),(\ref{2j}),
	(\ref{S}),(\ref{ee})},
\label{71b}
\end{align}
\end{subequations}
\end{spacing}
\noindent where $\textbf{D}_s^{(i)}$ is the obtained solution with $\gamma^{i}$ in the $i$-th iteration of the SROCR and $\textbf{u}_{\max}\big(\textbf{D}_s^{(i)}\big)$ is the eigenvector corresponding to $v_{\max}(\textbf{D}_s^{(i)})$, the largest eigenvalue of $\textbf{D}_s^{(i)}$. The parameter $\gamma^{(i)}\in[0,1]$ is introduced to manipulate the ratio of the largest eigenvalue of the obtained solution to the trace of $\textbf{D}_s$. Here, $\gamma^{(i)}=0$ means that the rank-one constraint is ignored directly, and $\gamma^{(i)}=1$ indicates that a rank-one solution can be found.
For a given  $\gamma^{i}$, problem~(\ref{PQES_SROCR}) is a convex problem and can be solved via CVX~\cite{CVX}. In this case, a local optimal rank-one solution can be acquired for problem~(\ref{PQES}) by iteratively increasing $\gamma^{i}$ from $0$ to $1$. 
Specifically, in each iteration, $\gamma^{i}$ is updated as
\vspace{-0.5 em}
\begin{spacing}{1.2}
\begin{equation}
\gamma^{i+1}=\min\bigg\{1,\frac{v_{\max}(\textbf{D}_s^{(i)})}{\mathrm{Tr}(\textbf{D}_s^{(i)})}+\delta^{(i)}\bigg\},
\end{equation}
\end{spacing}
\noindent where $\delta^{(i)}$ is the step size. 
If the pre-defined $\delta^{(i)}$ makes the problem infeasible, it can be reduced as $\delta^{(i+1)}=\delta^{(i)}/2$ until the problem is solvable~\cite{MuSROCR}. The termination condition for the SROCR algorithm is that $\mathrm{obj}(\textbf{D}_s^{(i)})-\mathrm{obj}(\textbf{D}_s^{(i-1)}) \le \epsilon_1$ an d $1-\gamma^{(i-1)}\le {\epsilon}_2$ are reached simultaneously,
where $\mathrm{obj}(\textbf{D}_s^{(i)})$ denotes the objective function value obtained with solution $\textbf{D}_s^{(i)}$, and $ {\epsilon}_1$ and $ {\epsilon}_2$ are the convergence thresholds. 

After the rank-one solution is found, we can get the passive beamforming coefficients $\{\textbf{d}_s\}$ via Cholesky decomposition as 
$\textbf{D}_s=\textbf{d}_s\textbf{d}_s^H, \forall s \in\{r,t\}.$

\begin{algorithm}[!t]
	\caption{QWSR-Optimal STAR-RIS assisted NOMA communications for ES}\label{alg}
	\begin{algorithmic}[1]
		\State Initialize a SIC decoding order, a preset maximal iteration number $l_{max}$, and a threshold $\epsilon$.
		\State Find the feasible solutions for $\{\textbf{w}_k^l\}$ and $\{\textbf{d}_s^l\}$ with $l=0$.
		\While {$l\leq l_{max}$ or the fractional increase of the objective value $\le \epsilon$}
		\State Obtain the optimal solution $\{\textbf{w}_k^{l+1}\}$ of~(\ref{PW_ESlow}) for a given $\{\textbf{d}_s^l\}$.
		\State Obtain the optimal solution $\{\textbf{d}_s^{l+1}\}$ of~(\ref{PQES}) for a given $\{\textbf{w}_k^{l+1}\}$.
		\State $l=l+1$.
		\EndWhile
	\end{algorithmic}
\end{algorithm}

\subsection{Complexity and Convergence Analysis}
\textcolor{black}{We summarize the overall algorithm in Algorithm~\ref{alg}, where the ABCs $\{\textbf{w}_k\}$ at the BS and the PTRCs $\{\textbf{d}_s\}$ at the STAR-RIS are optimized alternatively and iteratively. The obtained solutions in each iteration serve as the input local points of the next iteration until the algorithm termination conditions are satisfied.}
Explicitly, the problem complexity of the SDP subproblem for the ABCs design at the BS is \textcolor{black}{$\mathcal{O}\big(\max(5K^2+1,N)^4\sqrt{N}\log_2(1/\epsilon)\big)$} for a given solution accuracy $\epsilon>0$~\cite{Luo2010}. Additionally, The complexity of the subproblem for the PTRCs design is \textcolor{black}{$\mathcal{O}\big(l^{SRO}\max(5K^2-2K+5,M)^4\sqrt{M}\log_2(1/\epsilon)\big)$, where $l^{SRO}$ is the iteration numbers needed for the SROCR approach.}
In consequence, the total complexity of Algorithm~\ref{alg} is \textcolor{black}{$\mathcal{O}\Big(l^{max}\Big(\max\big(5K^2+1,N\big)^4\sqrt{N}+l^{SRO}\max(5K^2-2K+5,M)^4\sqrt{M}\Big)\log_2(1/\epsilon)\Big)$,} where $l^{max}$ is the actual maximum iteration numbers.

\begin{theorem}\label{theoremConvergence}
	The convergence of the proposed Algorithm~\ref{alg} is guaranteed.
\end{theorem}
\begin{IEEEproof}
	Please refer to Appendix~\ref{Proof_Convergence}.
\end{IEEEproof}

\section{Extension to the MS and TS Protocols}\label{sec_MSTSsolution}
In this section, we extend the proposed solution to the MS protocol and the TS protocol. For the MS protocol, we extend the algorithm for ES into a two-loop penalty-based iterative algorithm by adding an outer loop iteration controlled by the penalty factor. For the TS protocol, the optimization problem is decomposed into two subproblems, each of which can be solved with the method adopted for ES.

\subsection{Optimization for the MS protocol}

The main difference between the problems for the MS and ES protocols lies in that the continue amplitude $\beta_m^{s}\in[0,1],\forall s \in\{r, t\}$ in constraint~(\ref{1b}) becomes binary variable $\beta_m^{s}\in\{0,1\},\forall s \in\{r, t\}$. \textcolor{black}{Thus, the formulated problem~(\ref{PESMS}) for MS under a given decoding order $o_k$ is given by
\vspace{-0.8em}
\begin{spacing}{1}
\begin{subequations}\label{MS}
	\begin{align}
	\max _{\{\boldsymbol{\Theta}_s,\textbf{w}_k\}} \quad &  \sum_{k\in\mathcal{K}}Q_kR_{k} \label{P1MS_obj}\\
	\rm{s.t. }  \quad
	& \beta^t_m+\beta^r_m=1,
	\label{MS1b}
	\beta^r_m, \beta^t_m \in\{0, 1\},\\
	& \rm{(\ref{a}), (\ref{c})-(\ref{e})}.
	\label{MS1a}	
	\end{align}
\end{subequations}	
\end{spacing}
\noindent which can be solved following a similar process to that of the ES protocol by alternatively optimizing the ABCs $\{\textbf{w}_k\}$ and the PTRCs $\{\textbf{d}_s\}$. In this case, for the ABCs design at the BS under a given $\{\textbf{d}_s\},\forall s\in\{r,t\}$, we refer to the same method for the ES protocol to get the solution. In the following, we focus on the the PTRCs design at the STAR-RIS  under a given $\{\textbf{w}_k\},\forall k\in\mathcal{K}$.}

The constraint $\beta_m^{s}\in\{0,1\},\forall s \in\{r, t\}$ in~(\ref{MS1b}) is equal to equality constraint below
\vspace{-0.5em}
\begin{spacing}{1}
\begin{equation}\label{beta_equ}
\beta_m^{s}(1-\beta_m^{s})=0,\forall s \in\{r, t\}, m\in\mathcal{M}.
\end{equation}
\end{spacing}
\noindent Inspired by this, we can resort to a penalty-based algorithm proposed in~\cite{Mu2021} via adding the constraint~(\ref{beta_equ}) as a penalty term into the objective function. The optimization subproblem for the PTRCs design is expressed as
\vspace{-0.5 em}
\begin{spacing}{1}
\begin{subequations}\label{PMS}
\begin{align}
\max _{\substack{\{\textbf{D}_s,\textbf{W}_k,\boldsymbol{\beta}^s,\\R_{k},S_{kj},I_{kj}\}}} \quad &  \sum_{k\in\mathcal{K}}Q_kR_{k}-\eta\sum_{m\in\mathcal{M}}\sum_{s\in\mathcal\{r,t\}}\big(\beta_m^{s}(1-\beta_m^{s})\big)\label{PMS_obj}\\
\rm{s.t. }   \quad
& 
\rm{(\ref{2b})-(\ref{2e}),(\ref{2d})-(\ref{2h}),(\ref{2j})}.
\label{MSa}
\end{align}
\end{subequations}
\end{spacing}
\noindent where $\eta>0$ is the penalty factor for restricting the value of $\{\beta_m^s\}$ into the set $\{0,1\}$, otherwise the objective function will be penalized if $\{\beta_m^s\}$ falls into the range $(0,1)$.

However, the penalty term regarding the $\beta_m^s$  in the objective is non-convex. 
To deal with this challenge, we adopt the SCA approach to approximate this term by employing the first-order Taylor expansion. Particularly, for the given local point $\{\tilde{\textbf{D}}_s\}$, namely $\{\tilde{\boldsymbol{\beta}^s}\}$, an upper bound for the penalty term can be derived as
\vspace{-0.8em}
\begin{spacing}{1}
\begin{align}
\sum_{m\in\mathcal{M}}\big(\beta_m^{s}(1-\beta_m^{s})\big)\notag
&\le \sum_{m\in\mathcal{M}}\Big(\tilde{\beta}_m^s-(\tilde{\beta}_m^s)^2+(1-2\tilde{\beta}_m^s)({\beta}_m^s-\tilde{\beta}_m^s)\Big)\notag\\
&=\sum_{m\in\mathcal{M}}{\beta}_m^s-2\sum_{m\in\mathcal{M}}\tilde{\beta}_m^s{\beta}_m^s+\sum_{m\in\mathcal{M}}(\tilde{\beta}_m^s)^2 \notag\\
&=\mathrm{Tr}(\textbf{D}_s)-2\mathrm{Tr}(\tilde{\textbf{P}}_s\textbf{D}_s)+\mathrm{Tr}(\tilde{\textbf{P}}_s\tilde{\textbf{P}}_s) \\
&\triangleq \Lambda(\textbf{D}_s,\tilde{\textbf{D}}_s),\forall s\in\{r,t\},\notag
\end{align}
\end{spacing}
\noindent where $\tilde{\textbf{P}}_s$ is the corresponding diagonal matrix of $\tilde{\textbf{D}}_s$.
After that, the objective function in~(\ref{PMS}) is transformed as an affine function about the variables.

As for the non-convex constraint~(\ref{2e}), 
we can deal with them via a similar method as described for the ES protocol. After all the operations, this transformed problem is given as
\vspace{-0.5 em}
\begin{spacing}{1}
\begin{subequations}\label{PQMS}
\begin{align}
\max _{\{\textbf{D}_s,R_{k},S_{kj},I_{kj}\}} \quad &\sum_{k\in\mathcal{K}}Q_kR_{k} -\eta\sum_{s\in\mathcal\{r,t\}}\Lambda(\textbf{D}_s,\tilde{\textbf{D}}_s) \label{P8_obj}\\
\rm{s.t. }   \quad
&\rm{(\ref{2b}),(\ref{2c}),(\ref{2d}),(\ref{2h}), (\ref{2j}),
	(\ref{S}),(\ref{ee})},
\label{8a}
\end{align}
\end{subequations}
\end{spacing}
\noindent which is in the same form as problem~(\ref{PQES}) for ES, and can be solved via the same method. 

\begin{algorithm}[!t]
	\caption{QWSR-Optimal STAR-RIS assisted NOMA communications for MS}\label{algMS}
	\begin{algorithmic}[1]
		\State Initialize a SIC decoding order, a preset maximal iteration number $l_{max}$, the threshold $\tilde{\epsilon}_1$ and $\tilde{\epsilon}_2$, and the penalty factor $\eta$.
		\State Find the feasible solutions for $\{\textbf{w}_k^l\}$, $\{\textbf{d}_s^l\}$ with $l=0$.
	\Repeat:
		\While {$l\leq l_{max}$ or the fractional increase of the objective value $\le \tilde{\epsilon}_1$}
		\State Obtain the optimal solution $\{\textbf{w}_k^{l+1}\}$ of~(\ref{PW_ESlow}) in the MS version under the given $\{\textbf{d}_s^l\}$.
		\State Obtain the optimal solution $\{\textbf{d}_s^{l+1}\}$ of~(\ref{PQMS}) in the MS version under the given $\{\textbf{w}_k^{l+1}\}$.
		\State $l=l+1$.
		\EndWhile
		\State $\eta=\zeta \eta$.
	\Until the constraint violation is smaller than $\tilde{\epsilon}_2$.
	\end{algorithmic}
\end{algorithm}

Different from Algorithm~\ref{alg} for ES, a two-loop algorithm for MS is summarized in Algorithm~\ref{algMS}. 
In the outer loop, the penalty factor is gradually increased after each iteration, i.e., $\eta=\zeta\eta$ with $\zeta>1$.
The termination criterion in the outer loop is defined as
\vspace{-0.5 em}
\begin{spacing}{1.1}
\begin{equation}
\max\Big\{\beta_m^s-(\beta_m^s)^2,\forall s\in\{r,t\}, m\in\mathcal{M}\Big\}\le\tilde{\epsilon}_2,
\end{equation}
\end{spacing}
\noindent where $\tilde{\epsilon}_2$ is the predefined accuracy.
In the inner loop, $\{\textbf{W}_k,\textbf{D}_s\}$ are optimized alternately via the BCD method similar to the way for ES under the given penalty factor. Since the value of the objective function is non-decreasing in each iteration of the inner loop, the optimal value of the objective function is bounded below. Thus, this penalty-based iterative algorithm is guaranteed to convergence as the factor $\eta$ approaches infinity.

\subsection{Optimization for the TS protocol}

For a given $\alpha^{s},\forall s \in \{r,t\}$, this problem can be decomposed into two subproblem according to whether $k\in\mathcal{R}\text{ or }\mathcal{T}$ as 
\vspace{-0.8 em}
\begin{spacing}{1}
\begin{subequations}
\begin{align}
\max _{\{\boldsymbol{\Theta}_{s},\textbf{w}_k\}} \quad &  \sum_{k\in\mathcal{R}\text{ or }\mathcal{T},s\in\{r,t\}}Q_kR^{s}_{k} \label{Pts1_obj}\\
\rm{s.t. }   \quad
& \mathrm{{(\ref{a})-(\ref{e})}}.
\label{tsa1}
\end{align}
\end{subequations}
\end{spacing}
\noindent It is worth noting that each of these two subproblems can be solved like the problem~(\ref{PES}). Let $R^s_{sum},s\in\{r,t\}$ denote the maximal QWSR for the corresponding problem, then problem~(\ref{PESMS}) for TS is equivalent to the following problem
\vspace{-0.5 em}
\begin{spacing}{1}
\begin{subequations}
\begin{align}
\max _{\{\alpha^{s}\}} \quad &  \sum_{s\in\{r,t\}}\alpha^{s}R^{s}_{sum} \label{Pts2_obj}\\
\rm{s.t. }   \quad
& \mathrm{(\ref{alpha})}
\label{tsf2}
\end{align}
\end{subequations}
\end{spacing}
\noindent which is the linear programming and can be efficiently solved.
\textcolor{black}{
\begin{remark}
Based on the above analysis, it is worth noting that the optimization for the TS protocol in one time slot results in a per-slot single-surface working style of the STAR-RIS. However, as the data queues evolve with time, the working surface of the STAR-RIS will switch in different time slots to serve different user groups and keep all the queues stable from the perspective of a long term. This is because the employed time-varying queue length-based weights reflect the priority of each user at different moments.
\end{remark}}

\section{Simulation Results}\label{sec_simulation}
In this section, numerical results are provided to validate the effectiveness performance of our proposed algorithm.

\subsection{Simulation Configuration}

\begin{figure}[htb]
	\centering
	\includegraphics[width = 3.5 in]{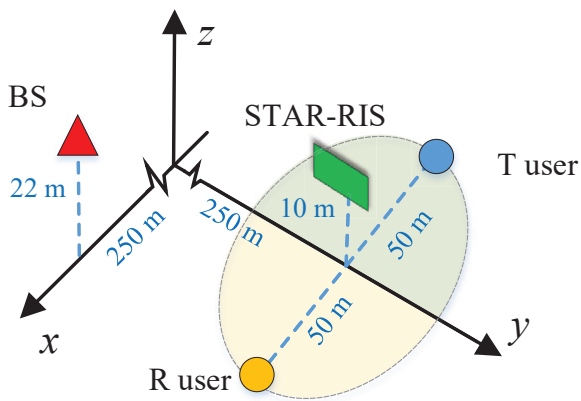}
	\caption{Simulation layout of the STAR-RIS assisted NOMA communication system.}
	\label{Fig_SimSetup}
\end{figure}

A shown in~Fig.~\ref{Fig_SimSetup}, a three-dimensional (3D) coordinate system is demonstrated to describe the locations of the transceivers in the STAR-RIS assisted NOMA communications. We assume that the BS is $22$ meters (m) tall and the START-RIS is located on a building with a height of $10$m, and they are set at the point ($250,0,22$) and the point ($0, 250, 10$), respectively. In addition, a transmission user and a reflection user are distributed on the opposite sides of the STAR-RIS with a horizontal distance from the STAR-RIS being $50$m, and their coordinates are ($-50, 250, 0$) and ($50, 250, 0$) in Fig.~\ref{Fig_SimSetup}.

In the simulation, the narrow-band quasi-static fading channels from the BS to the STAR-RIS and from the STAR-RIS to the users are modeled as Rician fading channels, shown as
\vspace{-0.8 em}
\begin{spacing}{1}
\begin{align}
\textbf{G}&=\sqrt{Pl(\rho_g)}\Bigg(\sqrt{\frac{\phi_g}{{\phi_g+1}}}\textbf{G}^{\rm{LOS}}+\sqrt{\frac{1}{{\phi_g+1}}}\textbf{G}^{\rm{NLOS}}\Bigg),\\
\textbf{v}_k&=\sqrt{Pl(\rho_v)}\Bigg(\sqrt{\frac{\phi_v}{{\phi_v+1}}}\textbf{v}_k^{\rm{LOS}}+\sqrt{\frac{1}{{\phi_v+1}}}\textbf{v}_k^{\rm{NLOS}}\Bigg),\forall k \in\mathcal{K},
\end{align}
\end{spacing}
\noindent where the two terms outside the brackets are the distance-dependent path loss, with $\rho_g$ and $\rho_v$ being the distance between the BS and the STAR-RIS and between the STAR-RIS and the users. Moreover, $\phi_g$ and $\phi_v$ are the Rician factors with $\phi_g = \phi_v=3$dB, and $\textbf{G}^{\rm{LOS}}$ and $\textbf{v}_k^{\rm{LOS}}$ are the corresponding deterministic LOS components, while $\textbf{G}^{\rm{NLOS}}$ and $\textbf{v}_k^{\rm{NLOS}}$ are the Rayleigh fading components.
According to the $3$rd Generation Partnership Project ($3$GPP) technical report for the urban macro (UMa) scenario~\cite{3gpp_channel}, the distance-dependent path loss in dB is given by
\vspace{-0.5 em}
\begin{spacing}{1}
\begin{equation}
Pl(\rho)=28+22\log_{10}(\rho)+20\log_{10}(f_c),
\end{equation}
\end{spacing}
\noindent
where $\rho$ is the distance between the transmitter and the receiver, and $f_c$ is the carrier frequency with $f_c=2$GHz. We assume that the BS is equipped with a uniform linear array and the STAR-RIS is equipped with a uniform planar array, with the antenna spacing of both arrays being half of the wavelength. Unless otherwise stated, the maximal transmit power at the BS is set as $P_{max}=40$W, with the signal-to-noise ratio (SNR) being $5$dB. The length of one slot is $1$ms. The data arrives randomly according to a Possion distribution with the average arrival rate being $\lambda_1=2$bits/s/Hz and $\lambda_2=6$bits/s/Hz. Besides, the antenna number of the BS is $N=4$ and the element number of the STAR-RIS is $M=20$. 

\begin{figure}[!t]
	\centering
	\begin{minipage}[t]{0.48\linewidth}
		\centering
		\includegraphics[height=2.3 in]{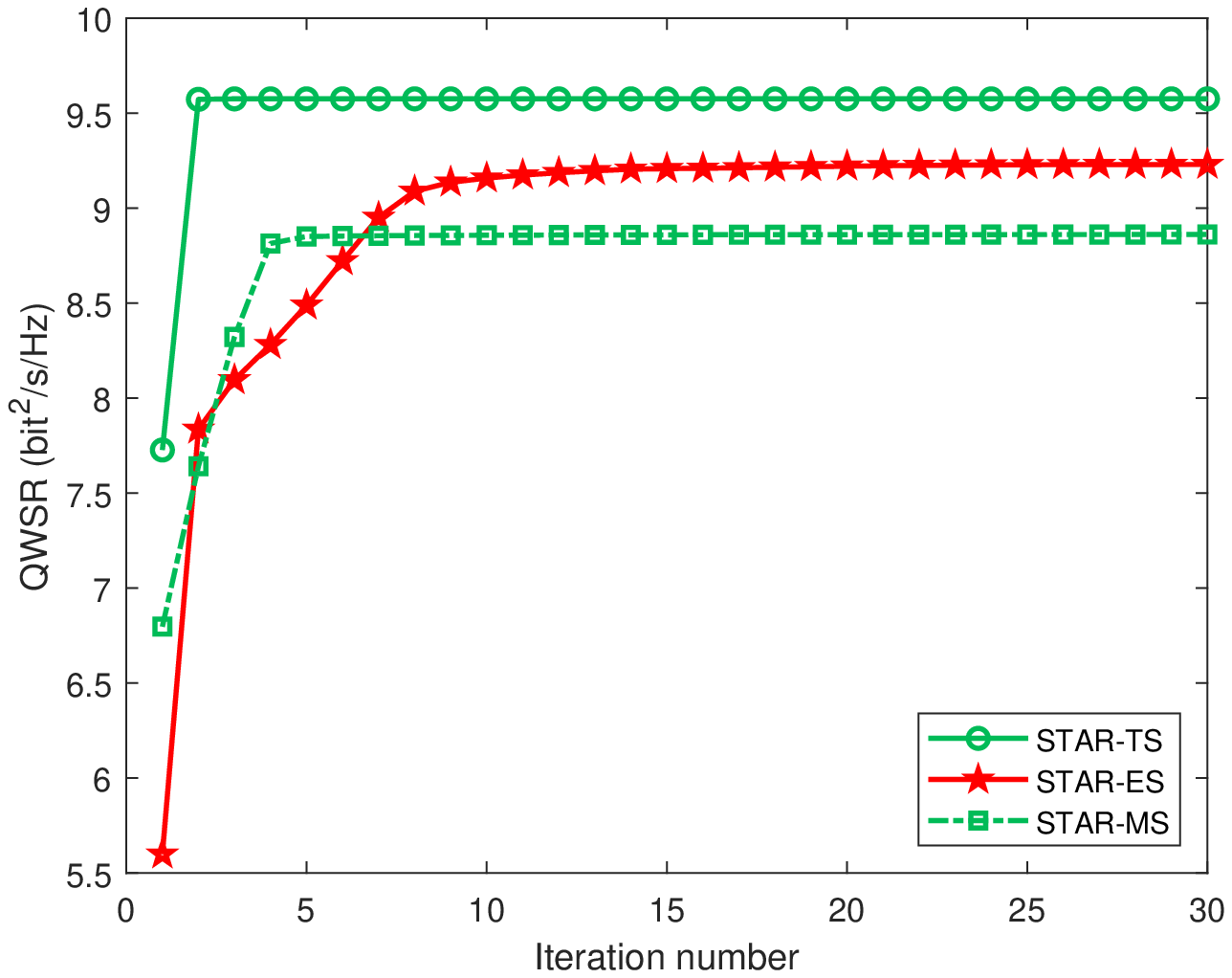}
		\caption{Convergence of the algorithms.}
		\label{Fig_Convergence}	
	\end{minipage}
	\begin{minipage}[t]{0.48\linewidth}
		\centering
		\includegraphics[height=2.3 in]{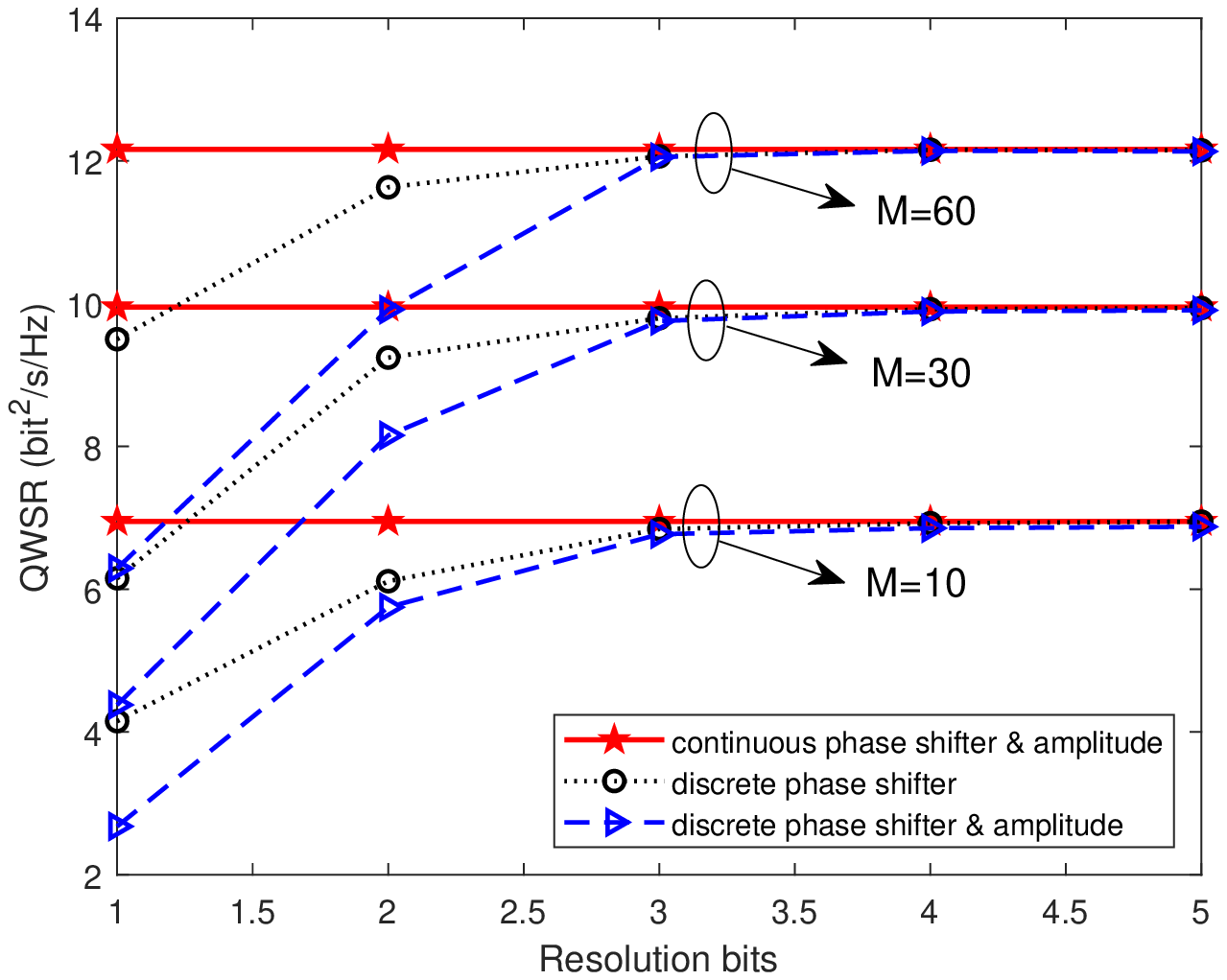}
		\caption{Quantization performance for ES.}
		\label{Fig_Resolution}		
	\end{minipage}
\end{figure}

\subsection{Convergence of the proposed Algorithms}

Fig.~\ref{Fig_Convergence} provides the convergence of our proposed algorithms by setting the threshold in Algorithm~\ref{alg} as $\epsilon=10^{-4}$. It can be seen from the figure that the algorithms converge approximately after $12$ iterations. Therefore, Theorem~\ref{theoremConvergence} is proven to be accurate. In addition, 
we can see that TS has superior QWSR performance than the others. The reason for this is that in TS, one group of users can utilize the full energy, but in ES and MS, the energy is split up by the STAR-RIS and used by two groups of users at the same time with mutual interference. That is, TS benefits from interference-free communications by inefficiently using the communication time. \textcolor{black}{However, due to the periodical switch-over of the elements, the time synchronization for TS involves a great level of complexity in hardware implementation.} 

\subsection{Quantization Performance}
Take the ES protocol as an example, Fig.~\ref{Fig_Resolution} shows the QWSR performance for different resolutions of the discrete phase shifter, as well as the both discrete phase shifter and discrete amplitude. Specifically, the uniform quantization and the discretization of one-half exponential powers are employed for the phase shifters and the amplitudes, respectively. From the figure we can see, compared with the discrete phase shifter, the performance of the both discrete phase shifter and discrete amplitude scheme deteriorates further with a small resolution bit. However, the performance gap between these two cases almost disappears when the resolution is equal to or greater than $3$ bits. Meanwhile, we notice that the performance loss between the discrete cases with a 3-bit resolution quantization and the continuous case can be negligible no matter what the value of the element number is.

\subsection{Baseline schemes}

To verify the good performance offered by integrating the STAR-RIS and NOMA, we denote our proposed algorithm as \textbf{STAR-ES/MS/TS} and adopt the following three baselines:
\begin{itemize}
	\item \textbf{Conventional RIS-assisted NOMA system (Conv-RIS)}: This scheme employs one reflecting-only RIS and one transmitting-only RIS, each
	consisting of $M/2$ elements to achieve full-space coverage for a fair comparison. This scheme can be regarded as a special case of the STAR-RIS, where half of the elements operate to reflect signals only and the other elements operate to transmit signals only. Thus, this problem can be solved by setting $\bm{\beta}^r=[\bm{1}_{1\times M/2}, \bm{0}_{1\times M/2}]$ and $\bm{\beta}^t=[\bm{0}_{1\times M/2}, \bm{1}_{1\times M/2}]$.
	\item \textbf{Uniform energy splitting (STAR-UES)}: In this case, the same amplitude coefficients (namely $\beta_m^s=0.5, \forall s \in \{r,t\}, m \in \mathcal{M}$)  are assumed to be employed among all elements of the STAR-RIS for transmission and reflection, respectively. It can be viewed as another special case of the STAR-RIS that utilizes a surface-by-surface design for amplitudes.
	\item \textbf{STAR-ES/MS/TS-OMA}: In this case, instead of NOMA, the BS transmit signals to the users through the orthogonal frequency/time resources~\cite{WuCoverage}. Let $\varpi_k\in[0,1]$ denote the proportion of resource blocks allocated to user $k$, which satisfies $\sum_{k\in \mathcal{K}}\varpi_k\le1$.
	Then, the achievable rate of user $k$ in this case is $R_k^O=\varpi_k\log2\big(1+\frac{|\textbf{v}_k\bm{\Theta}_{s_k}\textbf{G}\textbf{w}_k|^2}{\varpi_k\sigma^2}\big),\forall k\in\mathcal{K}$.	
\end{itemize}

\subsection{QWSR performance Comparison}
\begin{figure*}[htb]
	\centering
	\subfigure[QWSR versus different STAR-RIS element numbers.]{
		\label{Fig_EleNumber}
		\centering
		\includegraphics[height=2.3in]{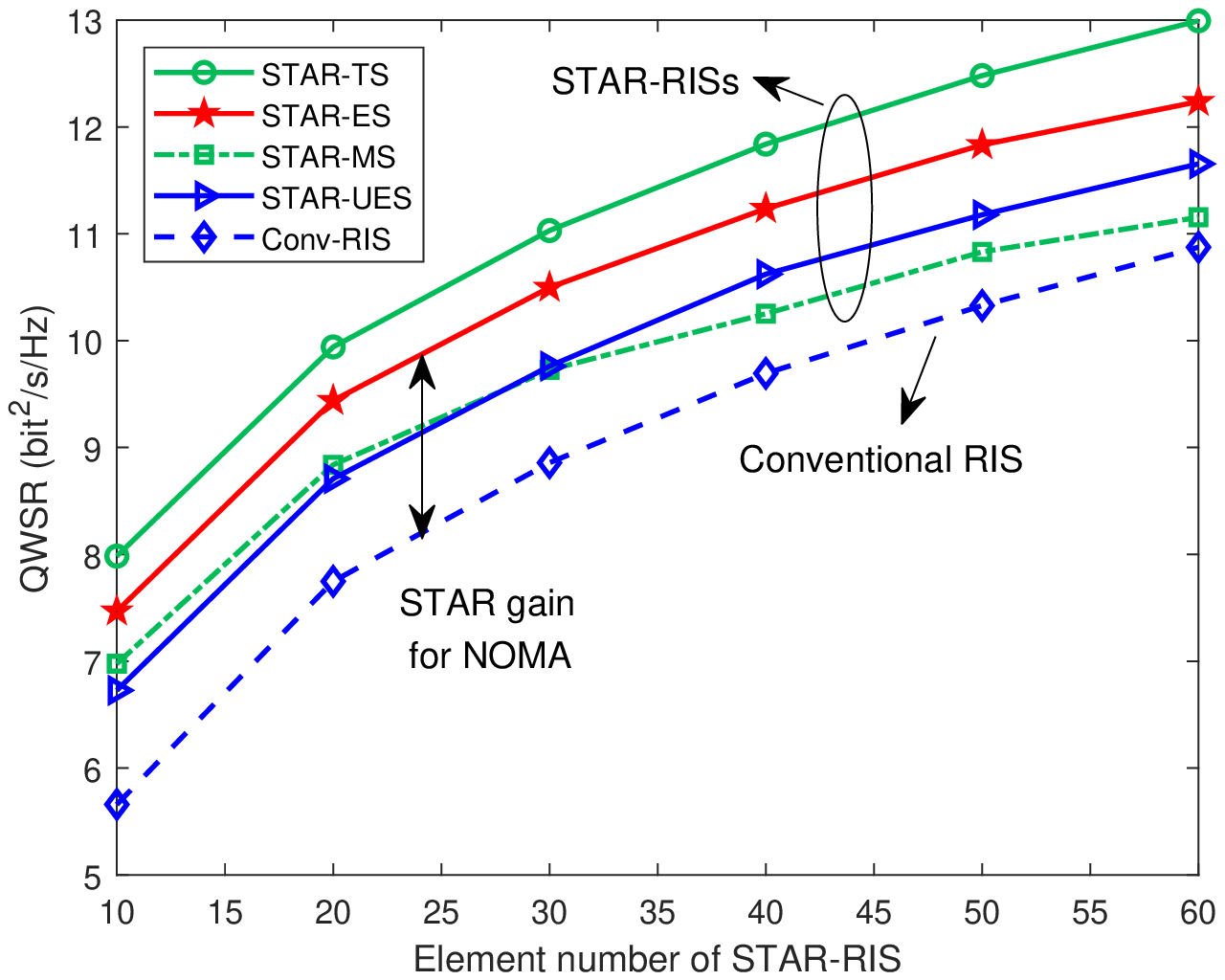}
	}	
	\subfigure[QWSR versus different SNR.]{
		\label{Fig_SNR}
		\centering						
		\includegraphics[height=2.3in]{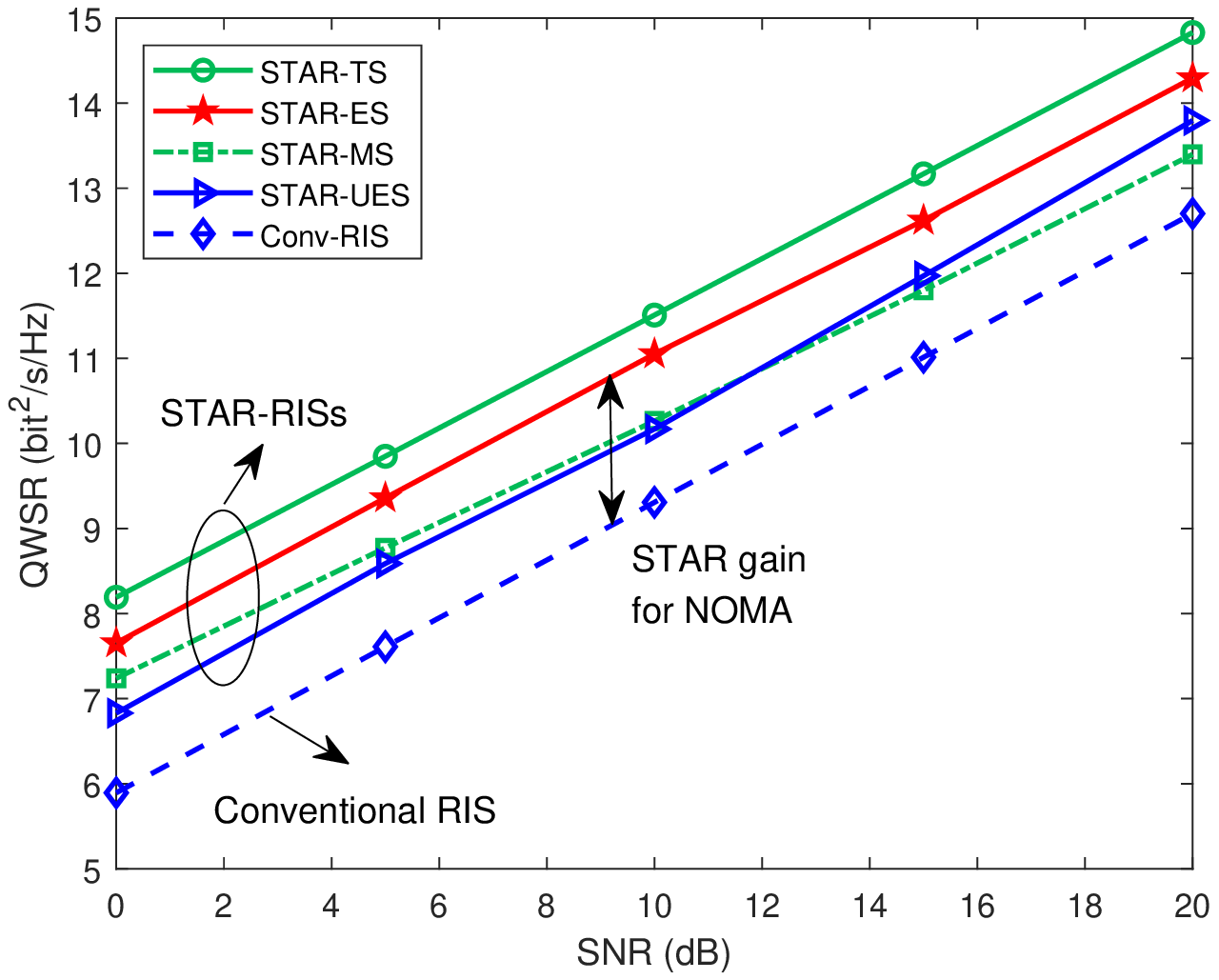}
	} 				
	\caption{QWSR comparison for the impact of STAR.}
	\label{Fig_QWSR_STAR}
\end{figure*}

\begin{figure*}[htb]
	\centering
	\subfigure[QWSR versus different STAR-RIS element numbers.]{
		\label{Fig_EleOMA}
		\centering
		\includegraphics[height=2.3in]{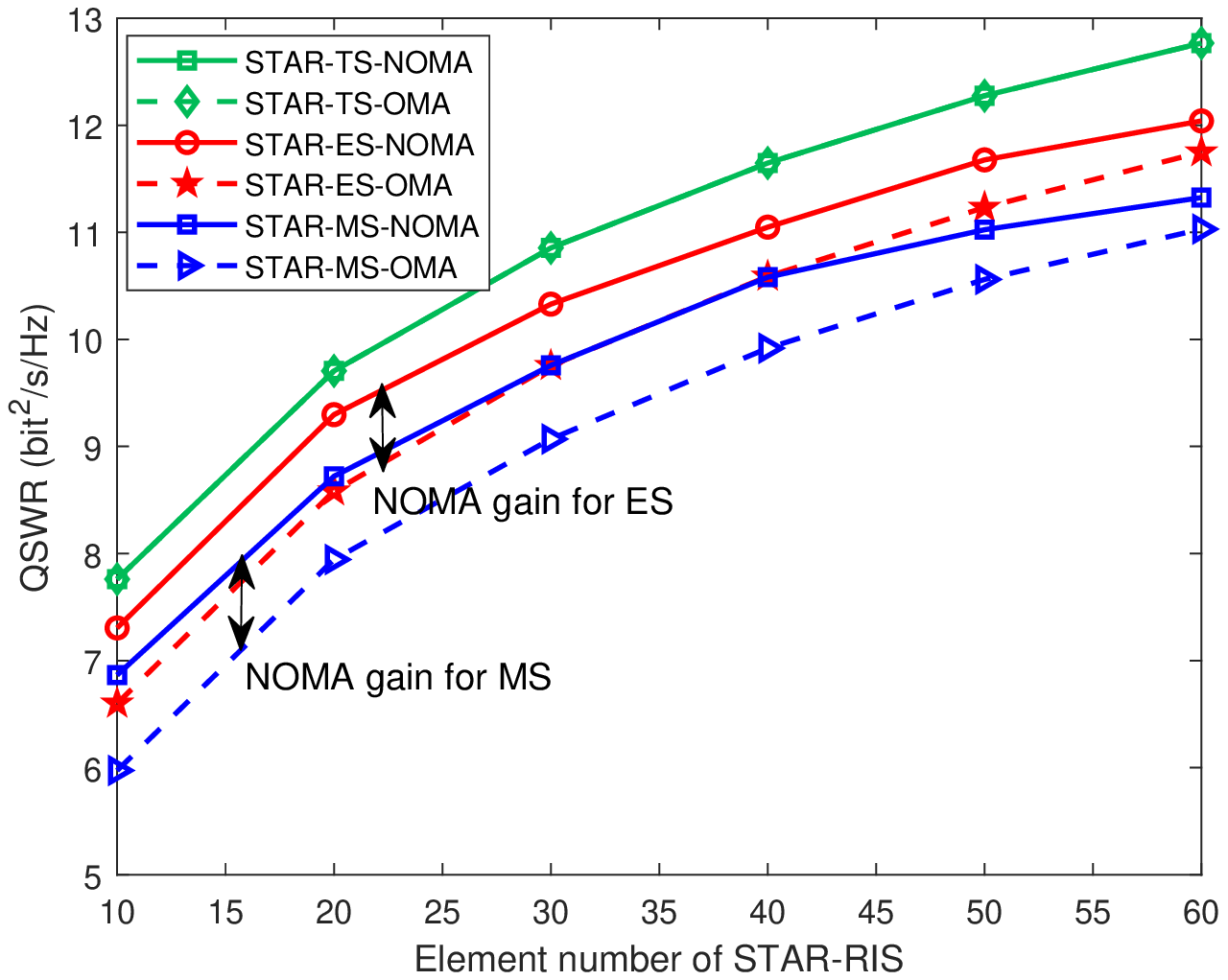}
	}	
	\subfigure[QWSR versus different SNR.]{
		\label{Fig_SNR_OMA}
		\centering						
		\includegraphics[height=2.3in]{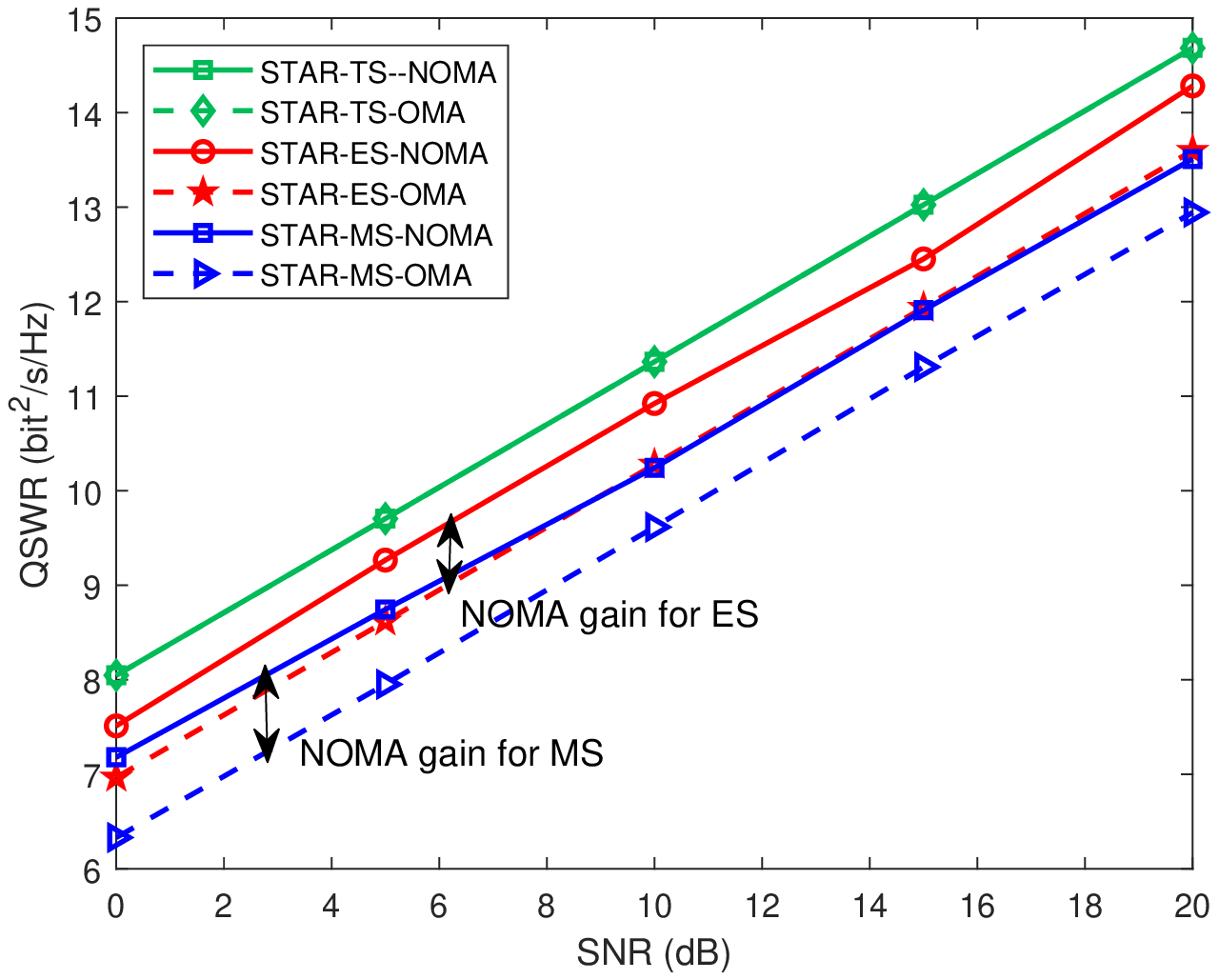}
	} 				
	\caption{QWSR comparison for the impact of NOMA.}
	\label{Fig_QWSR_OMA}
\end{figure*}

To demonstrate the performance of our proposed STAR-RIS assisted NOMA communications, we first compare QWSR performance in a single slot. In particular, Figs.~\ref{Fig_QWSR_STAR} reveals the impact of deploying STAR-RIS, and Fig.~\ref{Fig_QWSR_OMA} presents the impact of exploiting NOMA by comparing the corresponding OMA schemes.

Fig.~\ref{Fig_EleNumber} illustrates the QWSR performance versus different STAR-RIS element numbers.
The QWSR increases with the growing $M$ for the reason that more elements contribute to a higher array gain.
In addition, the STAR-RIS schemes achieve a significant performance gain compared with conventional RIS though NOMA is employed in both schemes, referred to as the ``STAR gain for NOMA", which verifies the superiority of the proposed STAR-RIS.
In addition, due to the binary value restrictions for the amplitude coefficients of transmission and reflection, MS suffers from some performance loss as compared to the ES and TS protocols. Nevertheless, MS is more attractive since that the on-off attribute of each element is easier to implement.

Fig.~\ref{Fig_SNR} compares the QWSR performance versus different SNR. As shown in the figure, the QWSR performance increases as the SNR grows. It can be observed that compared with the Conv-RIS scheme, a smaller SNR (i.e., less power) is needed for the STAR schemes to achieve the same QWSR. In addition, it should be highlighted that the unique characteristic of interference-free communications contributes to boosting the performance of TS no matter what the values of the SNR.

By comparing with OMA schemes, Fig.~\ref{Fig_QWSR_OMA} depicts the NOMA gain for STAR-RISs in terms of the QWSR performance. With STAR-RIS, the BS serves all users via the time-division multiple access under OMA. As observed from Fig.~\ref{Fig_EleOMA} and Fig.~\ref{Fig_SNR_OMA}, the NOMA schemes outperform the OMA schemes for both ES and MS. This is because the NOMA schemes allow all users to be served simultaneously under the ES and MS protocols, thus using the communication efficiently compared with the OMA schemes. Since the proposed TS protocol severs only one user in each time instant, there is no NOMA gain for TS in this two-user case. 

\begin{figure}[ht]
	\centering
	\includegraphics[height = 2.5 in]{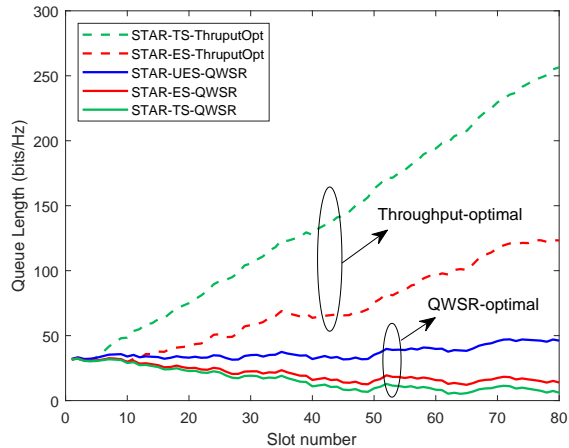}
	\caption{System stability}
	\label{Fig_emtStable}
\end{figure}

\subsection{Average Queue Length Comparison}

Fig.~\ref{Fig_emtStable} demonstrates the stability of the ES and TS protocols via comparing the average queue length over a long time with the schemes optimized by maximizing the sum rate, i.e., throughput-optimal cases without considering the queue-based weight. These cases are referred to as ``STAR-ES-ThruputOpt" and ``STAR-TS-ThruputOpt", respectively. Note that since MS is a special case of ES, we omit it here for conciseness. It can be observed that the throughput-optimal cases both for ES and TS have growing queues as time involves, which implies that the system cannot be stable since the average queue length is unbounded after an infinite horizon of time. This is indeed expected, since the inconsideration of the urgency of users results in an improper rate allocation, causing that queues of some users keep accumulating.
By contrast, the proposed QWSR-based schemes (i.e., ES, TS, and UES) achieve the state of dynamic equivalence, i.e., stable queues with limited oscillations, which proves that the reformulated QWSR maximization problem is able to guarantee the system stability in the long run.

\begin{figure*}[!t]
	\centering
	\subfigure[Average queue length versus different element numbers of the STAR-RIS.]{
		\label{Fig_emtEle}
		\centering
		\includegraphics[height=2.3in]{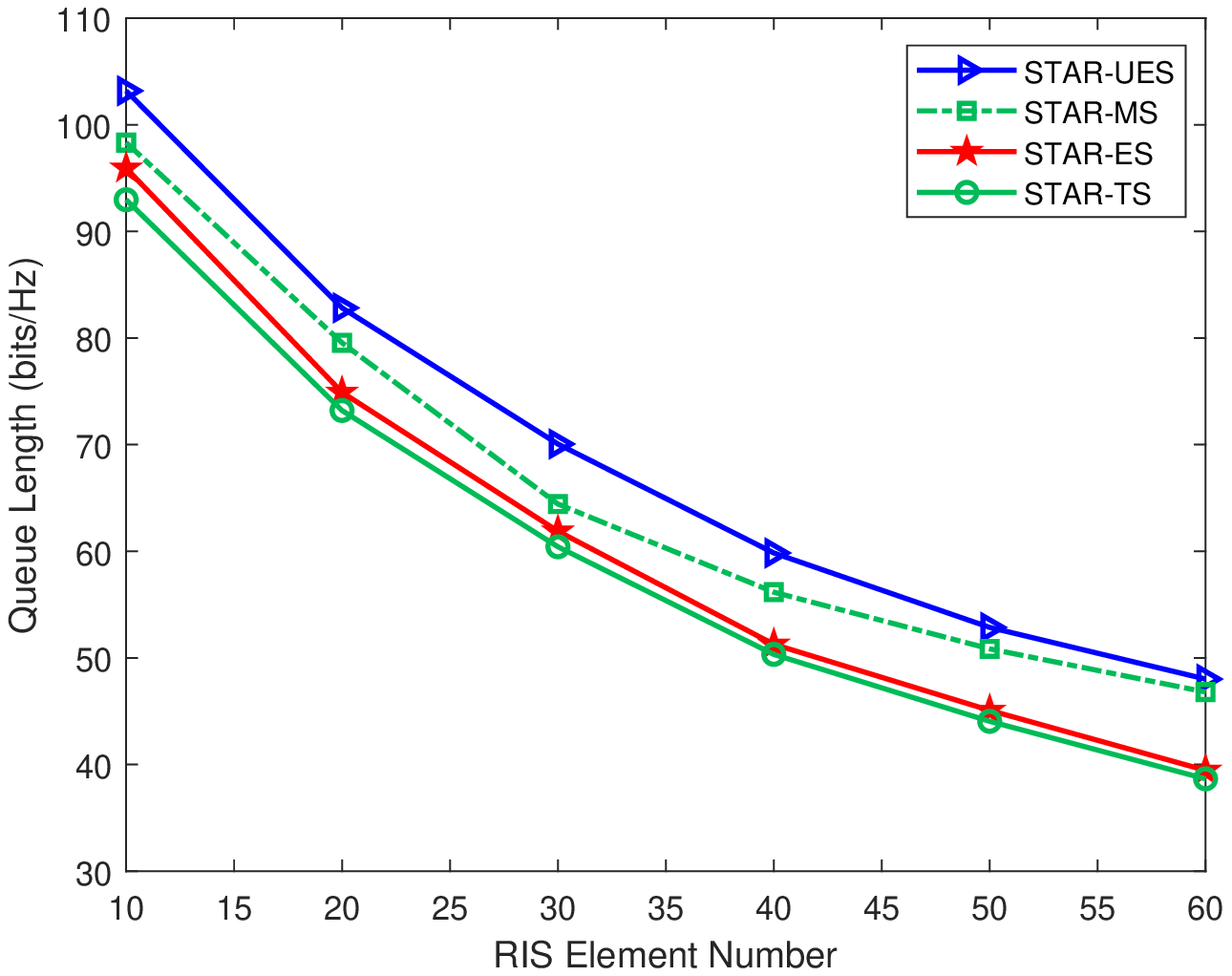}
	}	
	\subfigure[Average queue length versus different arrival rates.]{
		\label{Fig_emtArrival}
		\centering						
		\includegraphics[height=2.3in]{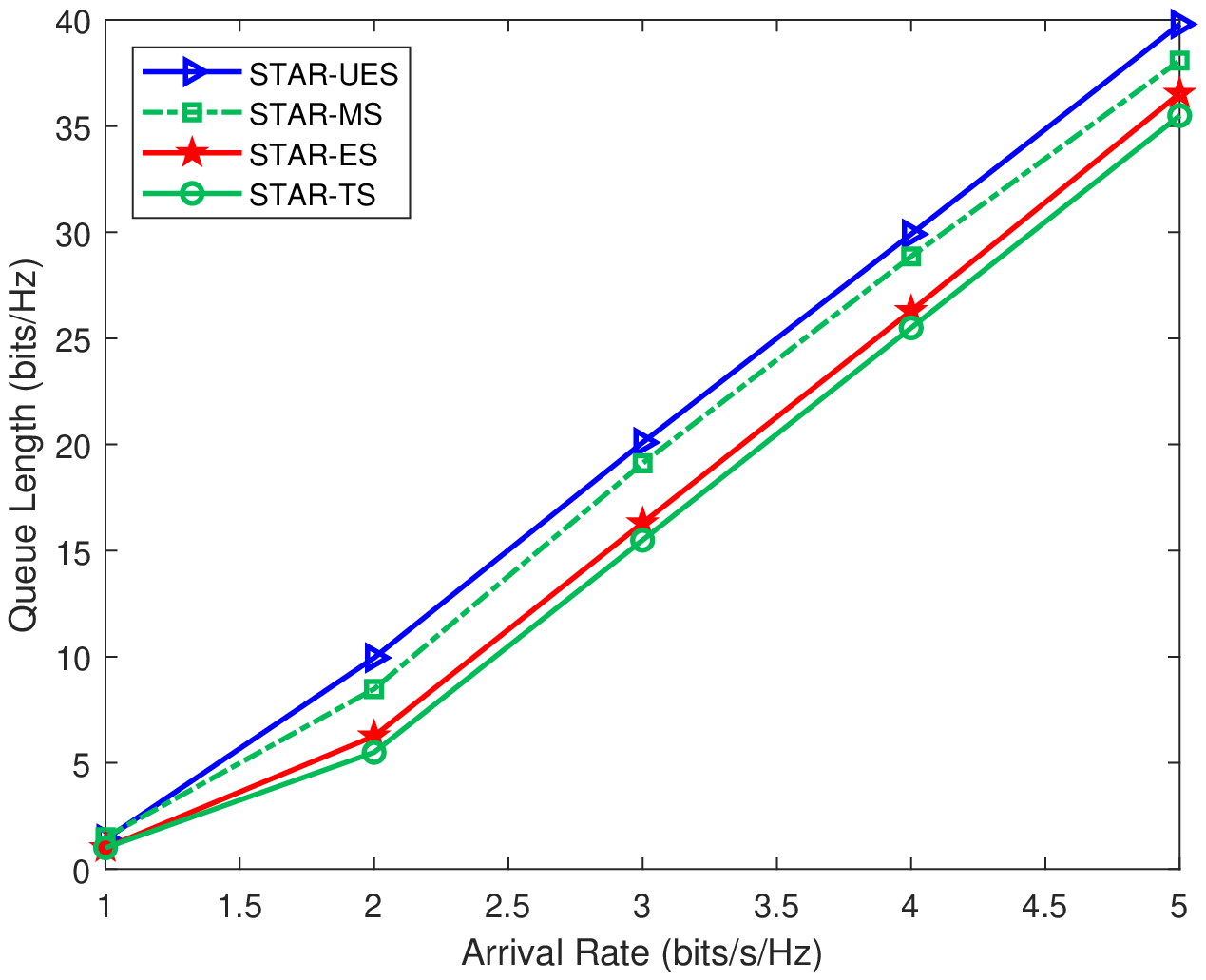}
	}
	\caption{Average queue length performance comparison.}
	\label{queue_length}
\end{figure*}

In Fig.~\ref{queue_length}, we compare the average queue length among the STAR-RIS schemes and the best performing baseline UES over $5$ slots.
Fig.~\ref{Fig_emtEle} compares the average queue length performance versus different element numbers of the STAR-RIS. It can be observed from the figure that a large element number leads to a lower queue length. This is reasonable since a large number of elements contributes to a higher beamforming gain and thus a higher rate, which in turn reduces the amount of data in the queue.

Fig.~\ref{Fig_emtArrival} demonstrates the average queue length performance versus different arrival rates of the data pending to be transmitted. Note that the performance gap almost disappears among all the algorithms when the arrival rate is very small, but becomes more noticeable with a higher rate. The reason for this trend is explained as follows. When the data arrives at a very low rate, fewer of them can be accumulated in the queue, and thus the average queue length is almost zero for all the algorithms. As the rate increases, the queue length grows accordingly due to the constrained transmission rate restricted by the limited energy budget. Accordingly, the advantages of the proposed algorithms show up.

\section{Conclusion}\label{sec_conclusion}

In this paper, the stability for the queue-aware STAR-RIS assisted NOMA communication system has been studied, \textcolor{black}{which was reformulated to maximize the per-slot QWSR of users. The employed rate weights were determined by the length of the data queues kept at the BS.
More explicitly, by jointly optimizing the NOMA decoding order, the ABCs at the BS, and the PTRCs at the STAR-RIS, three operating protocols for the STAR-RISs, including ES, MS, and TS, were considered.} 
\textcolor{black}{ For ES, the BCD and the SCA methods were exploited to iteratively handle the intrinsically coupled non-convex problem. 
Then for MS, the proposed iterative algorithm was further extended to exploit the penalty-based method. 
For TS, the formulated problem was decomposed into two subproblems, each of which can be handled in the same manner as introduced for ES.
Simulation results confirmed the queue stability under the reformulated QWSR maximization problem} and revealed that our proposed STAR-RIS assisted NOMA communication system achieves better performance compared with the conventional schemes. Furthermore, the simulations also showed that the TS protocol has superior performance among the three protocols in terms of both the QWSR and the average queue length.


\appendices{}

\section{\label{ProofTheom1} Proof of Theorem~\ref{Theom_RankW}}

In the absence of constraint~(\ref{2i}), the Lagrangian function of problem~(\ref{PQES}) is 
\vspace{-0.8 em}
\begin{spacing}{1.2}
	\begin{align}
	\mathcal{L}=
	&\sum_{k\in\mathcal{K}}\sum_{j:o_k\le o_j}a_{kj}\Big(\frac{1}{ S_{kj}}- \mathrm{Tr}(\textbf{W}_k\textbf{H}_{j}^H\textbf{D}_{s_j}\textbf{H}_{j})\Big)\notag\\
	&+\sum_{k\in\mathcal{K}}\sum_{j:o_k\le o_j}b_{kj}\Big(\sum_{i:o_k\le o_i}\mathrm{Tr}(\textbf{W}_{i}\textbf{H}_{j}^H\textbf{D}_{s_j}\textbf{H}_{j})+\sigma^2-I_{kj}\Big)	\\
	&+\sum_{i\in\mathcal{K}}\sum_{k\in\mathcal{K}}\sum_{j:o_k< o_j}x^i_{kj}\Big(
	\mathrm{Tr}(\textbf{W}_j\textbf{H}_{i}^H\textbf{D}_{s_i}\textbf{H}_{i})
	-\mathrm{Tr}(\textbf{W}_k\textbf{H}_{i}^H\textbf{D}_{s_i}\textbf{H}_{i})\Big)\notag\\
	&+ c\Big(
	\sum_{k\in\mathcal{K}} \mathrm{Tr}(\textbf{W}_{k})-P_{max}\big)-\sum_{k\in\mathcal{K}}\mathrm{Tr}(\textbf{W}_k\bm{\textbf{X}}_k)+L_0,\notag
	\end{align}
\end{spacing}
\noindent
	where $L_0$ is the terms independent of $\textbf{W}_k, k\in\mathcal{K}$. Moreover, the terms $a_{ki}$, $b_{kj}$, $c$, $x^i_{kj}$, and $\bm{\textbf{X}}_k$ are the Lagrange multipliers associated with the corresponding constraints. The Karush-Kuhn-Tucker (KKT) conditions for the optimal $\textbf{W}_k^*$ are displayed as follows
	\vspace{-0.5 em}
	\begin{spacing}{1.2}
	\begin{equation}\label{KKT}
	a_{kj}^*, b^{*}_{kj},c^*,x^{i^*}_{kj}\ge 0, \bm{\textbf{X}}_k\succeq\bm{0},
	\bm{\textbf{X}}_k^*\textbf{W}_k^*=\bm{0}, \nabla_{\textbf{W}_k^*}\mathcal{L}=\bm{0},
	\end{equation}
\end{spacing}
\noindent where $a_{kj}^*$, $b^{*}_{kj}$, $c^*$, $x^{i^*}_{kj}$, and $\bm{X}_k^*$ stand for the optimal Lagrange multipliers, and $\nabla_{\textbf{W}_k^*}\mathcal{L}$ is the gradients of $\mathcal{L}$ with respect to $\textbf{W}_k^*$. Then, we have the following equations,
\vspace{-0.5 em}
\begin{spacing}{1}
	\begin{align}\label{grad_L}
	\nabla_{\textbf{W}_k^*}\mathcal{L}=
	&-\sum_{j:o_k\le o_j}a_{kj}^*\big(\textbf{H}_{j}^H\textbf{D}_{s_j}\textbf{H}_{j}\big)^T
	+\sum_{m:o_m< o_k}\sum_{j:o_k\le o_j}b^*_{mj}\big(\textbf{H}_{j}^H\textbf{D}_{s_j}\textbf{H}_{j}\big)^T \notag \\
	&+\sum_{i\in\mathcal{K}}\Big(\sum_{m:o_m< o_k}x^{i^*}_{mk}
	-\sum_{j:o_k< o_j}x^{i^*}_{kj}\Big)\big(\textbf{H}_{i}^H\textbf{D}_{s_i}\textbf{H}_{i}\big)^T	+ c^*\textbf{I}-(\bm{\textbf{X}}^*_k)^T,\notag\\
	=&c^*\textbf{I}-(\bm{\textbf{X}}^*_k)^T-\sum_{i\in\mathcal{K}}y^k_i\big(\textbf{H}_{i}^H\textbf{D}_{s_i}\textbf{H}_{i}\big)^T=0,
	\end{align}
\end{spacing}
\noindent
with $y_i^k$ defined as
\vspace{-0.5 em}
\begin{spacing}{1.2}
\begin{equation}
y^k_i=\mathbbm{1}_{(o_k\le o_i)}a^*_{ki}-\mathbbm{1}_{(o_k\le o_i)}\sum_{m:o_m< o_k}b^*_{mi}-\Big(\sum_{m:o_m< o_k}x^{i^*}_{mk}
-\sum_{j:o_k< o_j}x^{i^*}_{kj}\Big),
\end{equation}
\end{spacing}
\noindent
where $\mathbbm{1}_{(\cdot)}$ is the indicator function, whose value is $1$ when the condition in $(\cdot)$ is true, and $0$ otherwise.

Let $\bm{\textbf{A}}_k=\sum_{i\in\mathcal{K}}y^k_i\big(\textbf{H}_{i}^H\textbf{D}_{s_i}\textbf{H}_{i}\big) $,  and rearrange the order of the items in formula~(\ref{grad_L}), we have
\begin{spacing}{1}
\begin{equation}\label{cA}
\bm{\textbf{X}}_k^*=c^*\textbf{I}-\bm{\textbf{A}}_k, \forall k \in\mathcal{K}.
\end{equation}
\end{spacing}
\noindent Let $z_k$ denote the maximal eigenvalue of $\bm{\textbf{A}}_k$ and $f_k$ be the algebraic multiplicity of the eigenvalue $z_k$. Reviewing~(\ref{KKT}), we know that $c^*\ge 0$ and $\bm{\textbf{X}}^*_k\succeq\bm{0}$. Next, we try to compare the value of $c^*$ and $z_k$. If $c^*<z_k$, it contradicts the condition $\bm{\textbf{X}}^*_k\succeq\bm{0}$. If $c^*>z_k$, the smallest eigenvalue of $\bm{\textbf{X}}_k^*$ is $c^*-z_k>0$, which indicate that $\bm{\textbf{X}}_k^*$ is a full rank positive-semidefinite matrix and its null space is zero, i.e., $\mathrm{Rank}(\textbf{W}_k)=0$ and $\textbf{W}_k$ is a zero matrix. However, considering the constraint $\sum_{k\in\mathcal{K}}\mathrm{Tr}(\textbf{W}_k) \le P_{max}$, zero matrix is not an effective solution for $\textbf{W}_k$. Therefore, $c^*$ can only be equal to $z_k$, i.e., $c^*=z_k $.

Since $z_k$ is the largest eigenvalue of $\bm{\textbf{A}}_k$, the other eigenvalues are less than $ c^*$. Combing equation~(\ref{cA}), we can know that $\bm{\textbf{X}}_k$ is a positive-semidefinite matrix with $f_k$ zero eigenvalues and $N-f_k$ positive eigenvalues, which means $\mathrm{Rank}(\bm{\textbf{X}}_k^*)=N-f_k$. Recalling that $\bm{X}_k^*\textbf{W}_k^*=\bm{0}$, the rank of $\textbf{W}_k^*$ satisfies 
\vspace{-0.5 em}
\begin{spacing}{1}
\begin{equation}
\mathrm{Rank}(\textbf{W}_k^*)\le\mathrm{Rank}(\text{the null space of } \bm{\textbf{X}}_k^*)=N-\mathrm{Rank}(\bm{\textbf{X}}_k^*)=f_k.
\end{equation}
\end{spacing}
\noindent In fact, due to the randomness of the channels, the probability that $\bm{\textbf{A}}_k$ has multiple eigenvalues with the same value (i.e., $f_k>1$) is very low. As such, we can say that $\mathrm{Rank}(\textbf{W}_k^*)=1$. This completes the proof.

\section{\label{Proof_Convergence} Proof of Theorem~\ref{theoremConvergence}}
Let $R_{sum}\big(\{\textbf{w}_k^l\}, \{\textbf{d}_s^l\}\big)$ denote the objective value of problem~(\ref{PES}) in the $l$-th iteration, then it follows that
\vspace{-1.3 em}
\begin{spacing}{1}
\begin{align}\label{Eq_proof2}
R_{sum}\big(\{\textbf{w}_k^l\}, \{\textbf{d}_s^l\}\big) &\overset{(a)}{\le} R_{sum}\big(\{\textbf{w}_k^{l+1}\}, \{\textbf{d}_s\}\big)
\overset{(b)}{\le} R_{sum}\big(\{\textbf{w}_k^{l+1}\}, \{\textbf{d}_s^{l+1}\}\big),
\end{align}
\end{spacing}
\noindent where, $(a)$ holds since that the optimal objective value for the ABC design we get is the lower bound of that of problem~(\ref{PW_ES}) for the given PTRCs $\{\textbf{d}_s^l\}$. Similarly, $(b)$ holds for the reason that the the optimal objective value for the PTRCs design serves as the lower bound of that of problem~(\ref{PQ_ES}) for the given ABC value $\{\textbf{w}_k^{l+1}\}$.

Moreover, it is suggested from Eq.~(\ref{Eq_proof2}) that the objective value of problem~(\ref{PES}) is no-decreasing after each iteration. Meanwhile, owing to the finite value that the system sum rate can achieve, the proposed algorithm is guaranteed to converge. 

\small
\begin{spacing}{1.1}

\end{spacing}

\end{document}